\documentclass[conference]{IEEEtran}
\IEEEoverridecommandlockouts
\usepackage{cite}
\usepackage{amsmath,amssymb,amsfonts}
\usepackage{algorithmic}
\usepackage{graphicx}
\usepackage{textcomp}
\usepackage{xcolor}
\usepackage{algorithm}
\usepackage{algorithmic}
\usepackage{fontspec}
\usepackage[caption=false,font=footnotesize]{subfig}

\makeatletter
\renewcommand{\ALG@name}{Algorithm}
\def\BibTeX{{\rm B\kern-.05em{\sc i\kern-.025em b}\kern-.08em
    T\kern-.1667em\lower.7ex\hbox{E}\kern-.125emX}}
    
\begin{document}

\title{Movable Antenna Assisted Dual-Polarized Multi-Cell Cooperative AirComp: An Alternating Optimization Approach
}

\author{
	\IEEEauthorblockN{
		Mingyu Hu\IEEEauthorrefmark{*},
		Nan Liu\IEEEauthorrefmark{+} and
		Wei Kang\IEEEauthorrefmark{*}
	\IEEEauthorblockA{\IEEEauthorrefmark{*}School of Information Science and Engineering, Southeast University, Nanjing, China 210096}
	\IEEEauthorblockA{\IEEEauthorrefmark{+}National Mobile Communications Research Laboratory, Southeast University, Nanjing, China 210096}
	\IEEEauthorblockA{\{myuhu, nanliu, wkang\}@seu.edu.cn}
}
\thanks{This work is partially supported by the National Natural Science Foundation of China under Grants $62361146853$, and $62371129$, and~the Research Fund of the National Mobile Communications Research Laboratory, Southeast University, under Grant 2025A05.}
}

\DeclareRobustCommand*{\IEEEauthorrefmark}[1]{
	\raisebox{0pt}[0pt][0pt]{\textsuperscript{\footnotesize\ensuremath{#1}}}}

\maketitle

\begin{abstract}
	Over-the-air computation  (AirComp) is a key enabler for distributed optimization, since it leverages analog waveform superposition to perform aggregation and thereby mitigates the communication bottleneck caused by iterative information exchange. However, AirComp is sensitive to wireless environment and conventional systems with fixed single-polarized base-station arrays cannot fully exploit spatial degrees of freedom while also suffering from polarization mismatch. To overcome these limitations, this paper proposes a multi-cell cooperative air-computation framework assisted by dual-polarized movable antennas (D-PMA), and formulates a mean squared error (MSE) minimization problem by jointly optimizing the combining matrix, polarization vectors, antenna positions, and user transmit coefficients. The resulting problem is highly nonconvex, so an alternating algorithm is developed in which closed-form updates are obtained for the combining matrix and transmit coefficients. Then a method based on successive convex approximation (SCA) and semidefinite relaxation (SDR) is proposed to refine polarization vectors, and the antenna positions are updated using a gradient-based method. In addition, we develop a statistical-channel-based scheme for optimizing the antenna locations, and we further present the corresponding algorithm to efficiently obtain the solution. Numerical results show that the proposed movable dual-polarized scheme consistently outperforms movable single-polarized and fixed-antenna baselines under both instantaneous and statistical channels.

\end{abstract}


\section{Introduction}
With the rapid evolution of wireless communication technologies, both the number of connected users and the volume of exchanged data have been growing explosively, posing significant challenges to physical-layer system design and the solution of related optimization problems. As a computing paradigm that decomposes a global function evaluation into computations performed at multiple subordinate nodes, distributed computing can effectively exploit local computational resources, thereby substantially reducing computational complexity and latency. Consequently, it has gradually emerged as a popular approach for addressing wireless communication problems \cite{tychogiorgos2013non}\cite{boukhedimi2017coordinated}, such as multi-cell cooperative computing architectures and cell-free massive MIMO communication systems. Taking a multi-cell cooperative system as an example, users first transmit their information to the associated base station (BS). After local processing at the BS, the processed information is forwarded to a central node. Such a hierarchical architecture avoids the need for a massive number of users to simultaneously connect to the central node, thereby alleviating backhaul congestion and reducing overall communication overhead. Moreover, distributed computing systems typically rely on iterative procedures that repeatedly update the current solution until the objective function converges. When cooperation is established among BS, the final iterates are guaranteed to reach consensus across BS, which in turn enhances the robustness of the system against channel variations and other sources of uncertainty. Meanwhile, a large body of literature has developed efficient distributed optimization algorithms, such as dual decomposition \cite{falsone2017dual} and the alternating direction method 
of multipliers (ADMM)\cite{boyd2011distributed,deng2017parallel,zhang2020proximal,chang2016asynchronous}, with theoretical guarantees on convergence.

With the rapid advancement of artificial intelligence (AI), distributed computing has become increasingly integrated with AI workloads, among which federated edge learning (FEEL) is a representative paradigm. Conventional model training typically relies on collecting data from multiple nodes into a centralized repository for storage and optimization. As the number of end users grows at massive scale, this centralized pipeline can incur substantial training complexity and storage overhead \cite{mcmahan2016federated} \cite{konevcny2016federated}. In contrast, FEEL trains models by exploiting local data and on-device computational resources at distributed clients, thereby significantly alleviating the computational burden on the central coordinator\cite{mcmahan2017communication}. Moreover, since raw user data are not required to be uploaded, FEEL can reduce the risk of privacy leakage arising from direct data exposure. Depending on whether data records and feature spaces overlap across participating clients, FEEL-based aggregation settings are commonly categorized into three classes: horizontal federated learning\cite{konevcny2016federated}, vertical federated learning\cite{du2004privacy}\cite{du2001privacy}, and federated transfer learning\cite{liu2020secure}\cite{zhang2022transfer}. Nevertheless, although FEEL avoids explicit data sharing through model aggregation, the exchange of model updates may still reveal sensitive information. Therefore, it is often necessary to protect transmitted model-related information. Homomorphic encryption (HE) enables learning or aggregation operations to be performed over encrypted values\cite{aono2017privacy,cheng2021secureboost,hardy2017private}; in a typical workflow, model updates are encrypted prior to transmission, and authorized parties that hold the corresponding secret key can decrypt the received information after communication. Differential privacy (DP), on the other hand, enhances confidentiality by perturbing transmitted statistics or updates with carefully calibrated noise\cite{bassily2014private}. However, such perturbation may degrade model utility, making the trade-off between privacy guarantees and predictive performance a central topic of ongoing research. In practical deployments, heterogeneous client capabilities can further challenge synchronous communication and training. To address these system-level constraints, asynchronous communication protocols\cite{duchi2013estimation}\cite{dai2015high} and client sampling strategies are widely adopted as effective solutions\cite{mcmahan2017communication}.

In the basic workflow of FEEL, the server first broadcasts an initial model to the participating clients. Each client then performs local updates of the model parameters using its on-device data and uploads the resulting model updates to the server. Upon reception, the server decodes the uploaded information from different clients and aggregates them to update the global model. This procedure is executed over multiple communication rounds, and when the number of clients is large, it can incur substantial communication overhead as well as considerable computational cost at the server\cite{chen2021distributed}. Over-the-air computation (AirComp) exploits the superposition property of wireless multiple-access channels\cite{zhu2021over}. Specifically, clients apply analog modulation to their transmitted signals such that, after simultaneous transmission, the signals naturally add up over the air and the desired function is computed implicitly in the analog domain. The server can then update the global model directly based on the aggregated result. As a consequence, the required radio resource blocks and the overall training latency can be significantly reduced, and the communication cost does not increase linearly with the number of participating clients. However, AirComp also injects channel impairments—such as fading and noise—into the aggregated outcome, which may degrade learning performance. To mitigate the resulting computation error, an aggregation filter can be employed at the server side to suppress distortion in the received superimposed signal\cite{lin2022distributed}.

Movable antenna (MA), as an emerging antenna paradigm, have been recognized as a promising enabler for future sixth-generation (6G) communication systems. Unlike conventional fixed antenna arrays, an MA system mounts radiating elements on structures that allow limited translation and/or rotation\cite{zhu2023movable,zhu2023modeling,ma2023mimo}. This mechanical or structural flexibility introduces additional spatial degrees of freedom, enabling the antenna geometry to adapt more effectively to the wireless propagation environment. Existing MA-array hardware implementations can be broadly categorized into several representative architectures, including mechanically actuated designs driven by motors or linear guides to reposition antenna elements\cite{ning2025movable}, liquid-metal based designs in which the radiating location is adjusted by redistributing conductive fluid within a container using external actuators\cite{morishita2013liquid,shen2021reconfigurable,wang2022continuous} and electronically reconfigurable antenna structures that alter the effective radiation configuration without relying solely on macroscopic mechanical displacement\cite{costantine2015reconfigurable} \cite{jiang2021pixel}. In most MA-enabled designs, antenna positions are determined by solving an optimization problem\cite{ma2024movable,shao20246d,shao20256d}. Typical objective functions include maximizing achievable communication rate and improving energy efficiency, while common constraints capture practical limits such as minimum inter-element spacing and bounded movement regions\cite{zhu2023beamforming}\cite{zhu2023movableantenna}. A growing body of studies has formulated scenario-specific optimization models and developed corresponding solution methods, with empirical and analytical evidence indicating that MA-based arrays can outperform traditional fixed arrays in terms of overall system performance under comparable conditions. In addition, when an electromagnetic (EM) wave propagates through a wireless environment, multipath propagation can distort its original polarization state. If the polarization of the receiving antenna is not aligned with that of the incident wave, a polarization mismatch loss will occur, leading to degraded received signal power and link quality. Compared with a conventional single-polarized antenna, a dual-polarized antenna integrates two mutually orthogonal polarization components, enabling the reception of EM waves with different polarization states. This capability provides additional degrees of freedom for system design—e.g., improved diversity, enhanced multiplexing opportunities, or more flexible multiple-access strategies—which can translate into better overall communication performance. Paper\cite{kwon2014optimal,nabar2002performance,kwon2014polarization,landon2007recovering} investigate the performance gains enabled by dual-polarized antenna techniques from different perspectives.

In this paper, a multi-cell distributed AirComp system assisted by dual-polarized MA is proposed. The main contributions are summarized as follows:

	\begin{itemize}
		\item To the best of our knowledge, although several existing studies have examined the performance gains brought by movable antennas in AirComp systems, this work is the first to investigate the additional benefits enabled by dual-polarized movable antenna (D-PMA). Moreover, we provide a systematic performance comparison against both conventional fixed positon antenna array (FPA) and MA.
		
		\item Prior studies have been largely limited to MA assisted single-base-station over-the-air computation AirComp scenarios. In contrast, this paper is the first to extend the investigation to a multi-cell setting. Specifically, we formulate an optimization problem that minimizes the sum of mean-squared error (MSE) across all base stations, where the decision variables include the polarization vectors, user transmit coefficients, antenna positions, and the combining beamforming matrices. The resulting design problem is highly non-convex, posing significant algorithmic challenges.
		
		\item This paper develops an efficient iterative optimization framework. Specifically, closed-form solutions are derived for the base-station combining beamforming matrices and the user transmit coefficients. For the polarization vectors at the user side and the base-station side, we propose two dedicated iterative methods based on successive convex approximation (SCA) and semidefinite relaxation (SDR), respectively. Finally, for antenna-position design, a gradient-descent based optimization procedure is introduced to update the movable-antenna locations.
		
		\item Numerical results demonstrate that the proposed algorithm converges within a finite number of iterations. In addition, we examine how various system parameters affect the MSE performance, and the results indicate that the proposed dual-polarized movable-antenna architecture achieves a smaller computation error than both single-polarized movable-antenna arrays and conventional fixed antenna arrays.

	\end{itemize}
	
	The remainder of this paper is organized as follows. The system model and the formulation of the optimization problem are presented in Section II. The proposed algorithm is described in Section III. Section IV investigates the optimization problem under statistical channel state information (CSI) and develops the corresponding solution method. Numerical results are provided in Section V. Section VI concludes the paper. Appendices A and B present the proofs of Lemmas 1 and 2, respectively.

\begin{figure}[t]
	\centering
	\vspace{1mm} 
	\includegraphics[width=0.9\columnwidth, keepaspectratio, trim=0 0 0 0, clip]{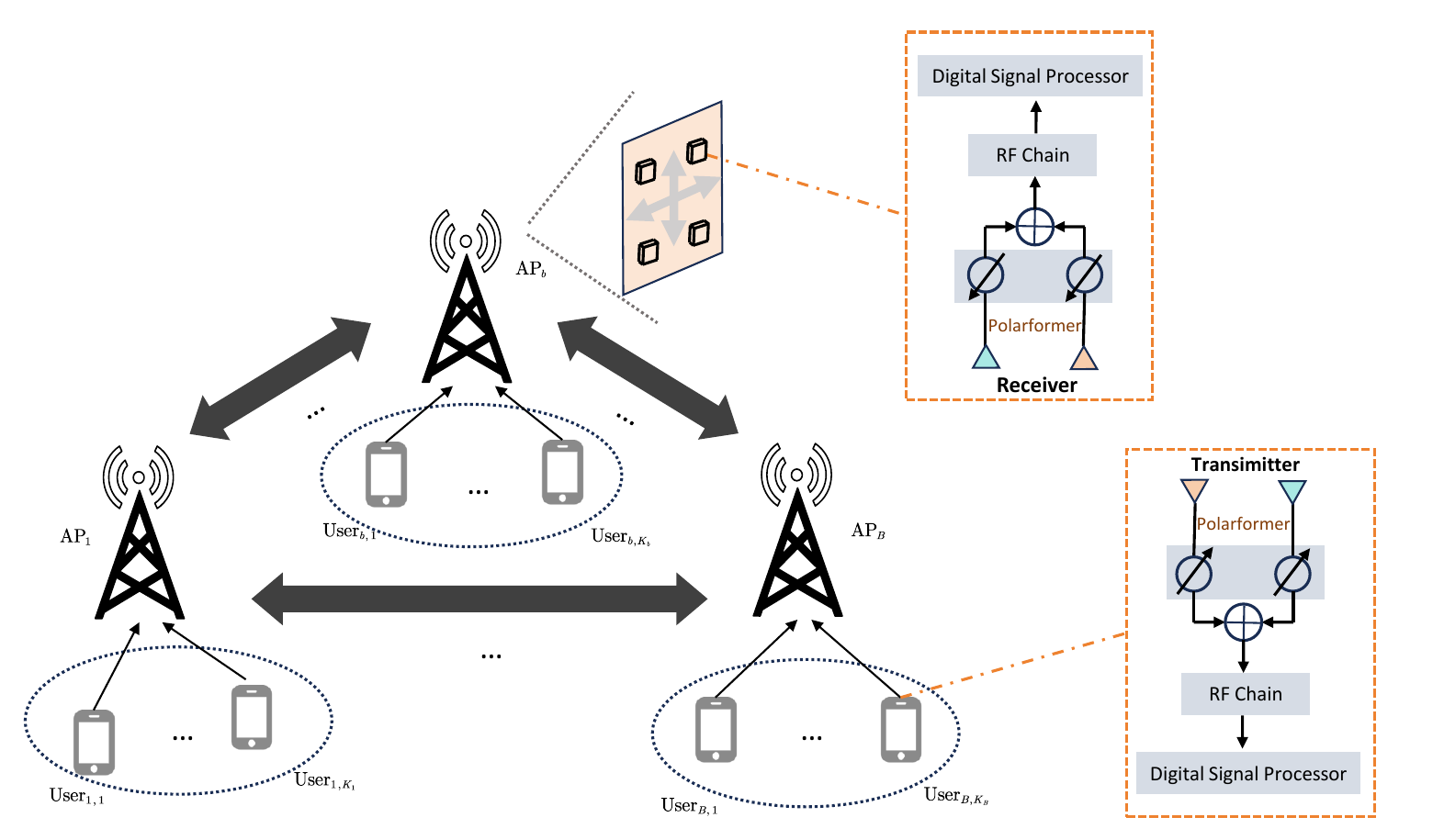}
	\caption{System model of D-PMA-aided AirComp system.}
	\label{fig:system_model}
	\vspace{-2mm} 
\end{figure}

\section{System Model}
	In this paper, we consider a D-PMA aided multi-cell cooperative AirComp system, as illustrated in Fig.~\ref{fig:system_model}. The network consists of $B$ cells. In each cell, one BS serves $K$ single-antenna users. All users in the same cell simultaneously transmit their local signals to the associated BS over the uplink. After receiving the superimposed signals, the BSs exchange processed information among one another through backhaul links to enable cooperative AirComp across cells.
	
	Each BS is equipped with a two-dimensional planar MA array comprising $M$ antennas. Each movable antenna element is connected to the BS controller via a flexible cable, which allows the antenna to be repositioned within a predefined rectangular region on the plane, subject to mechanical and wiring constraints. By appropriately adjusting the horizontal and vertical positions of the MA elements, the small-scale fading of both inter-cell and intra-cell wireless links can be effectively shaped, thereby providing additional spatial degrees of freedom for performance optimization.
	
	Furthermore, each transmit/receive antenna element is implemented as a dual-polarized unit consisting of two orthogonal linear polarization elements. One element is responsible for transmitting/receiving the horizontally polarized signal, while the other element handles the vertically polarized signal. For each polarization element, both the amplitude and the phase of the RF signal can be flexibly adjusted, enabling fine-grained adaptation to the surrounding wireless environment and enhancing the capability of the BSs to exploit polarization diversity in the considered cooperative AirComp system.
	
	\subsection{Channel Model}
	In this section, a channel model based on the field-response representation is introduced\cite{ma2023mimo}. Specifically, the inter-base-station  links are characterized by channel matrices, whereas the user-to-base-station links are modeled by channel vectors. 
	Subsequently, this model is generalized to account for the deployment of dual-polarized antenna elements.

	\begin{figure}[t]
		\centering
		\vspace{1mm} 
		\includegraphics[width=0.9\columnwidth, keepaspectratio, trim=0 0 0 0, clip]{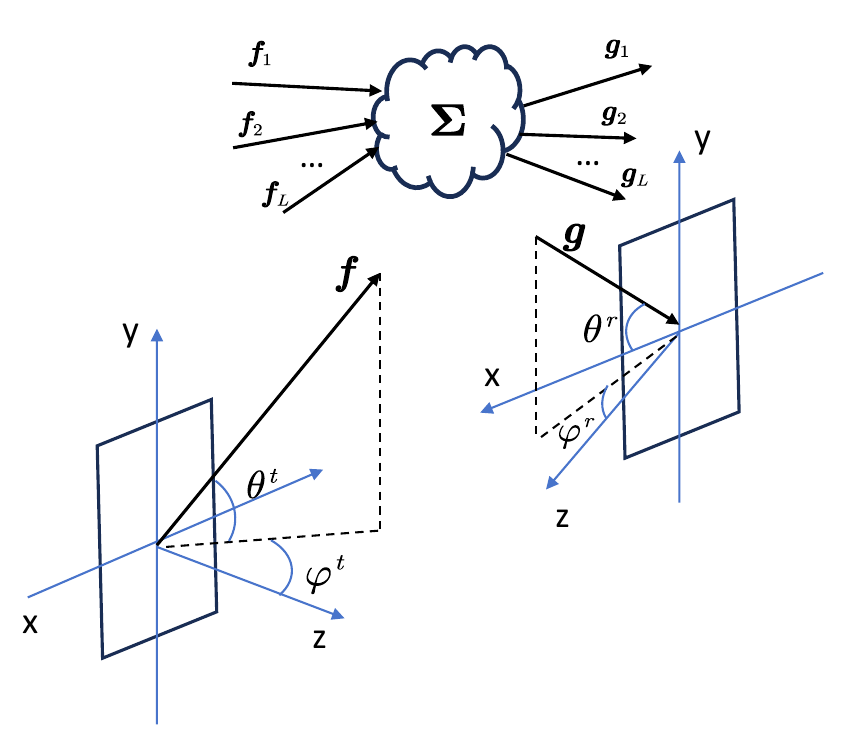}
		\caption{Description of signal propagation angles and scattering environment}
		\label{fig:CCS}
		\vspace{-2mm} 
	\end{figure}

	\subsubsection{Responsive Channel Model}
		First, we present the field-response channel model without employing polarized antennas. The field response model in \cite{ma2023mimo} is adopted for channel modeling. Let the set of BSs be denoted by $\mathcal{B} = \{1,2,\ldots,B\}$, the set of users in each cell by $\mathcal{K} = \{1,2,\ldots,K\}$, and the set of MA elements at each BS by $\mathcal{M} = \{1,2,\ldots,M\}$. 
		To facilitate the computation of the distances between different BSs as well as those between the users and their serving BSs, a global Cartesian coordinate system (CCS) is established. 
		Moreover, in order to conveniently describe the directions of the incident and radiated signals, a local CCS is introduced for each BS, whose origin is chosen as the geometric center of its planar MA array. 
		The local position of the $m$-th antenna element at BS $i$ is then represented by the vector $\mathbf{r}_{i,m} = (x_{i,m}, y_{i,m})^{\mathrm T}$, $i \in \mathcal{B}$, $m \in \mathcal{M}$. Define $\tilde{\mathbf{r}_i} = (\mathbf{r}_{i,1}^{\mathrm T},\cdots,\mathbf{r}_{i,M}^{\mathrm T})^{\mathrm T} \in \mathbb{R}^{2M}, \forall i \in \mathcal{B}$ as the antenna position vector of BS $i$. We assume that the number of propagation paths from each user to its serving BS, as well as the number of transmit/receive propagation paths between any pair of BSs, is identical and denoted by $L$. Let the set of propagation paths be denoted by $\mathcal{L} = \{1,2,\ldots,L\}$. On this basis, we first characterize the uplink channel from user $k$ to its associated BS $i$ by the channel vector $\mathbf{h}_{i,k} \in \mathbb{C}^{M \times 1}$. Let $\mathbf{u}_{i,k} \in \mathbb{C}^{L \times 1}$ denote the vector of complex fading coefficients associated with the $L$ propagation paths from user $k$ to BS $i$. The receive field-response matrix from user $k$ to BS $i$ is defined as
		\begin{equation}
			\mathbf{Q}_{i,k} =
			\begin{bmatrix}
				e^{j \frac{2\pi}{\lambda} \mathbf{v}_{i,k,1}^{\mathrm T} \mathbf{r}_{i,1}} & \cdots & e^{j \frac{2\pi}{\lambda} \mathbf{v}_{i,k,1}^{\mathrm T} \mathbf{r}_{i,M}} \\
				\vdots & \ddots & \vdots \\
				e^{j \frac{2\pi}{\lambda} \mathbf{v}_{i,k,L}^{\mathrm T} \mathbf{r}_{i,1}} & \cdots & e^{j \frac{2\pi}{\lambda} \mathbf{v}_{i,k,L}^{\mathrm T} \mathbf{r}_{i,M}}
			\end{bmatrix}
			\in \mathbb{C}^{L \times M},
			\label{response1}
		\end{equation}
		where $\lambda$ denotes the carrier wavelength, the vector $\mathbf{v}_{i,k,\ell}
		=
		\bigl(
		\cos\theta_{i,k,\ell}^{u}\sin\varphi_{i,k,\ell}^{u},
		\ \sin\theta_{i,k,\ell}^{u}
		\bigr)^{\mathrm T}$ denotes the unit-norm direction vector of the $\ell$-th propagation path from user $k$ to BS $i$,
		where $\theta_{i,k,\ell}^{u} \in [-\frac{\pi}{2},\frac{\pi}{2}]$ and $\varphi_{i,k,\ell}^{u} \in [-\pi,\pi]$ denote, respectively, the elevation angle of arrival and the azimuth angle of arrival of the incident uplink signal associated with the $\ell$-th path. Therefore, the uplink channel vector from user $k$ to BS $i$ can be expressed as
		\begin{equation}
			\mathbf{h}_{i,k}^{gen} = \mathbf{Q}_{i,k}^{\mathrm H}\mathbf{u}_{i,k} \in \mathbb{C}^{M \times 1},
			\label{hik}
		\end{equation}
		
		Similarly, the transmit and receive field-response matrices from BS $j$ to BS $i$ are respectively defined as
		\begin{equation}
			\mathbf{F}_{i,j} =
			\begin{bmatrix}
				e^{j\frac{2\pi}{\lambda}\mathbf{f}_{i,j,1}^{\mathrm T}\mathbf{r}_{j,1}} & \cdots & e^{j\frac{2\pi}{\lambda}\mathbf{f}_{i,j,1}^{\mathrm T}\mathbf{r}_{j,M}} \\
				\vdots & \ddots & \vdots \\
				e^{j\frac{2\pi}{\lambda}\mathbf{f}_{i,j,L}^{\mathrm T}\mathbf{r}_{j,1}} & \cdots & e^{j\frac{2\pi}{\lambda}\mathbf{f}_{i,j,L}^{\mathrm T}\mathbf{r}_{j,M}}
			\end{bmatrix}
			\in \mathbb{C}^{L\times M},
			\label{Fij}
		\end{equation}
		\begin{equation}
			\mathbf{G}_{i,j} =
			\begin{bmatrix}
				e^{j\frac{2\pi}{\lambda}\mathbf{g}_{i,j,1}^{\mathrm T}\mathbf{r}_{i,1}} & \cdots & e^{j\frac{2\pi}{\lambda}\mathbf{g}_{i,j,1}^{\mathrm T}\mathbf{r}_{i,M}} \\
				\vdots & \ddots & \vdots \\
				e^{j\frac{2\pi}{\lambda}\mathbf{g}_{i,j,L}^{\mathrm T}\mathbf{r}_{i,1}} & \cdots & e^{j\frac{2\pi}{\lambda}\mathbf{g}_{i,j,L}^{\mathrm T}\mathbf{r}_{i,M}}
			\end{bmatrix}
			\in \mathbb{C}^{L\times M},
			\label{Gij}
		\end{equation}
		where, $\mathbf{f}_{i,j,\ell} = (\cos\theta_{i,j,\ell}^{t}\sin\varphi_{i,j,\ell}^{t},\, \sin\theta_{i,j,\ell}^{t})^{\mathrm T}$ and $\mathbf{g}_{i,j,\ell} = (\cos\theta_{i,j,\ell}^{r}\sin\varphi_{i,j,\ell}^{r},\, \sin\theta_{i,j,\ell}^{r})^{\mathrm T}$ are respectively the transmit and receive direction vectors of the $l$-th path between BS $j$ and BS $i$, where $\theta_{i,j,\ell}^{t} \in [-\frac{\pi}{2},\frac{\pi}{2}]$ and $\varphi_{i,j,\ell}^{t} \in [-\pi,\pi]$ denote the elevation and azimuth angles of the transmit signal, and $\theta_{i,j,\ell}^{r} \in [-\frac{\pi}{2},\frac{\pi}{2}]$ and $\varphi_{i,j,\ell}^{r} \in [-\pi,\pi]$ denote the elevation and azimuth angles of the receive signal.  Fig.~2 illustrates the scattering environment and the Cartesian coordinate systems for signal transmission between base stations. Furthermore, the channel matrix from BS $j$ to BS $i$ is given by:
		\begin{equation}
			(\mathbf{H}_{i,j}^{gen})^{\mathrm{H}}
			= \mathbf{G}_{i,j}^{\mathrm H}\boldsymbol{\Sigma}_{i,j}\mathbf{F}_{i,j}
			\in \mathbb{C}^{M\times M},
			\label{Hijgen}
		\end{equation}
		where $\boldsymbol{\Sigma}_{i,j}$ is the channel fading matrix between BS $j$ and BS $i$, which is a diagonal matrix whose diagonal elements represent the fading coefficients of the corresponding propagation paths.
		
		\subsubsection{Polarized Channel Model}
		Next, we present the polarized channel model with the incorporation of polarized antennas. As shown in Fig.~1, each polarized antenna consists of two orthogonal polarization elements, one for horizontal polarization and the other for vertical polarization. By cooperating with each other, these two elements can transmit and receive electromagnetic waves more efficiently than a single-polarized antenna. Each polarization element is connected to a phase shifter, which can independently adjust the amplitude and phase of the electromagnetic wave, thereby flexibly controlling the overall polarization state of the antenna so that it matches the polarization direction of the electromagnetic wave as closely as possible. Although each base station is equipped with more than one antenna, all antennas are located within the same planar array and therefore experience the same wireless environment. Consequently, all MAs share an identical receive polarization vector. Specifically, the receive polarization vector of BS $i$ is given by
		\begin{equation}
			\boldsymbol{\varpi}_i = 
			\bigl[e^{j\psi_{i,1}},\,
			e^{j\psi_{i,2}}\bigr]^{\mathrm T}, i\in \mathcal{B},
			\label{varpi}
		\end{equation}
		where
		, $\psi_{i,c} \in [-\pi,\pi]$, $c \in \{1,2\}$, represents the phase shift of the $c$-th polarization element at cell $i$. The transmit polarization vector of the $k$-th user belonging to BS $i$ is
		\begin{equation}
			\mathbf{m}_{i,k}
			=
			\bigl[e^{j\psi_{i,k,1}},\,e^{j\psi_{i,k,2}}\bigr]^{\mathrm T},
			\quad i \in \mathcal{B},~k \in \mathcal{K},
			\label{mik}
		\end{equation}
		where $\psi_{i,k,c} \in [-\pi,\pi]$, $c \in \{1,2\}$, is the phase shift of the $c$-th polarization element of the $k$-th user under BS $i$. Assume that the horizontal element of the D-PMA is placed along the $x$-axis and the vertical element is placed along the $y$-axis. Therefore, two unit direction vectors along the corresponding directions can be defined as $\mathbf{e}_v = [0,1,0]^{\mathrm T}$ and $\mathbf{e}_h = [1,0,0]^{\mathrm T}$. The polarization state of an electromagnetic wave can be characterized by two orthogonal electric–field components. We select the following two sets of orthogonal unit vectors to represent the two orthogonal transmit/receive electric-field components for user-BS channel\cite{castellanos2023linear}:
		\begin{equation}
			\begin{aligned}
				\mathbf{z}_{i,k}^u &=
				\big[\sin\theta_{i,k}^u\sin\varphi_{i,k}^u,-\cos\theta_{i,k}^u,\sin\theta_{i,k}^u\cos\varphi_{i,k}^u\big]^{\mathrm{T}},\\
				\bar{\mathbf{z}}_{i,k}^u &=
				\big[\cos\varphi_{i,k}^u,0,-\sin\varphi_{i,k}^u\big]^{\mathrm{T}}, 
			\end{aligned}
			\label{zik}
		\end{equation}
		where the superscript $u$ denotes the association between the BS and the user,
		$\theta_{i,k}^u = \frac{1}{L}\sum_{\ell=1}^{L}\theta_{i,k,\ell}^{u}$ and $\varphi_{i,k}^{u} =\frac{1}{L} \sum_{\ell=1}^{L}\varphi_{i,k,\ell}^{u}$. To capture the characteristics of the electromagnetic wave on the two orthogonal polarization elements, we introduce the polarization field response matrix, whose entries are given by the projections of the two electric-field components onto the two orthogonal polarization directions. The transmit polarization field response matrix from user $k$ to base station $i$ is therefore given by
		\begin{equation}
			\mathbf{P}_{i,k}^u =
			\begin{bmatrix}
				\mathbf{z}_{i,k}^{u}\cdot\mathbf{e}_{v} & \mathbf{z}_{i,k}^{u}\cdot\mathbf{e}_{h}\\
				\bar{\mathbf{z}}_{i,k}^{u}\cdot\mathbf{e}_{v} & \bar{\mathbf{z}}_{i,k}^{u}\cdot\mathbf{e}_{h}
			\end{bmatrix} \in \mathbb{C}^{2 \times 2},
			\label{Pik}
		\end{equation}
		where $\cdot$ denotes the dot product. Similarly, the receive polarization field response matrix from user $k$ to base station $i$ is given by
		\begin{equation}
			\mathbf{Q}_{i,k}^u =
			\begin{bmatrix}
				\mathbf{e}_{v}\cdot\mathbf{z}_{i,k}^{u} & \mathbf{e}_{v}\cdot\bar{\mathbf{z}}_{i,k}^{u} \\
				\mathbf{e}_{h}\cdot\mathbf{z}_{i,k}^{u} & \mathbf{e}_{h}\cdot\bar{\mathbf{z}}_{i,k}^{u}
			\end{bmatrix} \in \mathbb{C}^{2 \times 2}
			\label{Qik}
		\end{equation}
		
		Consequently, the overall polarization field response matrix from user $k$ to base station $i$ is given by
		\begin{equation}
			\mathbf{A}_{i,k}^u = \mathbf{Q}_{i,k}^u\mathbf{P}_{i,k}^u \in \mathbb{C}^{2 \times 2}.
			\label{Aik}
		\end{equation}
		
		Each element of this matrix represents the complex field response from the user's vertical/horizontal polarization element to the BS's vertical/horizontal polarization element, respectively. Accordingly, all possible polarization channel state information between the vertical/horizontal ports of user $k$ and those of base station $i$ can be expressed as
		\begin{equation}
			\bar{\mathbf{h}}_{i,k}
			= \mathbf{h}_{i,k}^{\mathrm{gen}} \otimes \mathbf{A}_{i,k}^u
			\in \mathbb{C}^{2M \times 2}.
			\label{barhik}
		\end{equation}
		
		By introducing the receive polarization vector at the base station and the transmit polarization vector at the user side, the complete polarized channel from user $k$ to base station $i$ can be expressed as
		\begin{subequations}
			\begin{align}
				\mathbf{h}_{i,k}
				&= \left(\mathbf{I} \otimes \boldsymbol{\varpi}_{i}^{\mathrm{H}}\right)\bar{\mathbf{h}}_{i,k}\mathbf{m}_{i,k} \label{13a}\\
				&= \left(\mathbf{I} \otimes \boldsymbol{\varpi}_{i}^{\mathrm{H}}\right)
				\left(\mathbf{h}_{i,k}^{\mathrm{gen}} \otimes \mathbf{A}_{i,k}^u\right)\mathbf{m}_{i,k}\label{13b} \\
				&\overset{(a)}{=} 
				\underbrace{\mathbf{h}_{i,k}^{\mathrm{gen}}}_{\text{unpolarformed channel}} \times
				\underbrace{\boldsymbol{\varpi}_{i}^{\mathrm{H}}\mathbf{A}_{i,k}^u\mathbf{m}_{i,k}}_{\text{polarformed channel}}
				\in \mathbb{C}^{M \times 1}.\label{13c}
			\end{align}
		\end{subequations}
		where (a) follows from the property of the Kronecker product, i.e., 
		$(\mathbf{A} \otimes \mathbf{B})(\mathbf{C} \otimes \mathbf{D}) = (\mathbf{A}\mathbf{C}) \otimes (\mathbf{B}\mathbf{D})$. Similarly, the polarization channel matrix between base stations can be obtained. Specifically, the orthogonal electric-field components of the link from base station $j$ to base station $i$ can be represented by the following orthonormal unit vectors:
		\begin{equation}
			\begin{aligned}
				\mathbf{z}_{i,j}^{c} &= 
				\big[\sin\theta_{i,j}^{c}\sin\varphi_{i,j}^{c},\,-\cos\theta_{i,j}^{c},\,\sin\theta_{i,j}^{c}\cos\varphi_{i,j}^{c}\big]^{\mathrm{T}},\\
				\bar{\mathbf{z}}_{i,j}^{c} &= 
				\big[\cos\varphi_{i,j}^{c},\,0,\,-\sin\varphi_{i,j}^{c}\big]^{\mathrm{T}}, \quad c \in \{t,r\},
			\end{aligned}
			\label{zij}
		\end{equation}
		where $\theta_{i,j}^{c} = \frac{1}{L}\sum_{\ell=1}^{L}\theta_{i,j,\ell}^{c},
		\varphi_{i,j}^{c} = \frac{1}{L}\sum_{\ell=1}^{L}\varphi_{i,j,\ell}^{c}$. The superscript $\{t, r\}$ denotes transmit and receive between BSs respectively. Accordingly, the transmit and receive polarization field response matrices from base station $j$ to base station $i$ are given, respectively, by
		\begin{align}
			\mathbf{P}_{i,j}^t &=
			\begin{bmatrix}
				\mathbf{z}_{i,j}^{t}\cdot\mathbf{e}_{v} & \mathbf{z}_{i,j}^{t}\cdot\mathbf{e}_{h}\\
				\bar{\mathbf{z}}_{i,j}^{t}\cdot\mathbf{e}_{v} & \bar{\mathbf{z}}_{i,j}^{t}\cdot\mathbf{e}_{h}
			\end{bmatrix} \in \mathbb{C}^{2 \times 2}, \label{Pij}\\
			\mathbf{Q}_{i,j}^r &=
			\begin{bmatrix}
				\mathbf{e}_{v}\cdot\mathbf{z}_{i,j}^{r} & \mathbf{e}_{v}\cdot\bar{\mathbf{z}}_{i,j}^{r} \\
				\mathbf{e}_{h}\cdot\mathbf{z}_{i,j}^{r} & \mathbf{e}_{h}\cdot\bar{\mathbf{z}}_{i,j}^{r}
			\end{bmatrix} \in \mathbb{C}^{2 \times 2}. 
			\label{Qij}
		\end{align}
		
		Consequently, all possible polarization channel state information between base station $j$ and base station $i$ can be expressed as
		\begin{equation}
			\bar{\mathbf{H}}_{i,j}
			= \mathbf{H}_{i,j}^{\mathrm{gen}} \otimes \mathbf{A}_{i,j}^b
			\in \mathbb{C}^{2M \times 2M}.
			\label{barHij}
		\end{equation}
		where $\mathbf{A}_{i,j}^b = \mathbf{Q}_{i,j}^b\mathbf{P}_{i,j}^b \in \mathbb{C}^{2 \times 2}$ is the overall polarization feild response matrix from BS j to i. The superscript $b$ denotes  the association between BSs. Therefore, the polarization channel matrix from base station $j$ to base station $i$ is given by
		\begin{subequations}
			\begin{align}
			\mathbf{H}_{i,j}
		    &	= \left(\mathbf{I} \otimes \boldsymbol{\varpi}_{i}^{\mathrm{H}}\right)
			\left(\mathbf{H}_{i,j}^{\mathrm{gen}} \otimes \mathbf{A}_{i,j}^b\right)
			\left(\mathbf{I} \otimes \boldsymbol{\varpi}_{j}\right)\label{18a}\\
			&=\mathbf{H}_{i,j}^{\mathrm{gen}} \times \boldsymbol{\varpi}_{i}^{\mathrm{H}} \mathbf{A}_{i,j}^b \boldsymbol{\varpi}_{j} \in \mathbb{C}^{M \times M}.\label{18b}
		\end{align}
		\label{Hij}
		\end{subequations}
		Note that when $j=i$, $\mathbf{H}_{i,j}=\mathbf{I}$, where $\mathbf{I}$ is the identity matrix. In particular, when the number of propagation paths satisfies $L \gg 1$, the elevation and azimuth angles are assumed to be independently and symmetrically uniformly distributed around the origin. As a result, their sample means tend to zero, and the polarization field response matrices $\mathbf{A}_{i,j}^b$ and $\mathbf{A}_{i,k}^u$ asymptotically degenerate to identity matrices, i.e., $\mathbf{A}_{i,j}^b \approx \mathbf{I}$ and $\mathbf{A}_{i,k}^u \approx \mathbf{I}$. In this regime, the polarized channels reduce to $\mathbf{h}_{i,k} \approx \mathbf{h}_{i,k}^{\mathrm{gen}} \times \big(\boldsymbol{\varpi}_{i}^{\mathrm{H}}\mathbf{m}_{i,k}\big)$ and $\mathbf{H}_{i,j} \approx \mathbf{H}_{i,j}^{\mathrm{gen}} \times \big(\boldsymbol{\varpi}_{i}^{\mathrm{H}}\boldsymbol{\varpi}_{j}\big)$.

	\subsection{Signal Model}
	As illustrated in Fig.~1, the users in each cell transmit their signals to the serving base station, after which the base stations exchange information with one another, i.e., they acquire global channel state information in a cooperative manner. In this paper, since the number of users is sufficiently large and the transmit power is small, we model the inter-cell interference and noise as a Gaussian random variable.
	In the following, we will present the received signal model at each BS.
	
	The transmit symbol vector of the users in cell $i$ can be defined as
	$\mathbf{s}_{i} = [s_{i,1},\ldots,s_{i,K}]^{\mathrm{T}} \in \mathbb{C}^{K}$, 
	$\forall i \in \mathcal{B}$. We assume that the data symbols are zero-mean and normalized, i.e.,
	$\mathbb{E}\{s_{i,k}\} = 0$ and $\mathbb{E}\{|s_{i,k}|^{2}\} = 1$, 
	$\forall i \in \mathcal{B},\, k \in \mathcal{K}$. Moreover, the symbols are mutually independent across different cells and users, i.e., 
	$\mathbb{E}\{s_{i,k} s_{j,m}^{\mathrm{H}}\} = 0$, 
	$\forall i,j \in \mathcal{B},\, k,m \in \mathcal{K}$ with $i \neq j$ or $k \neq m$.
	where  The corresponding transmit coefficient vector of the $K$ users in cell $i$ is denoted by
	$\mathbf{a}_{i} = [a_{i,1},\ldots,a_{i,K}]^{\mathrm{T}} \in \mathbb{C}^{K}, \forall i \in \mathcal{B}$. Therefore, the received signal at base station $i$ can be written as
	\begin{equation}
		\mathbf{z}_{i}
		= \sum_{k=1}^{K} \mathbf{h}_{i,k} a_{i,k} s_{i,k} + \mathbf{n}_{i},
		\label{zi}
	\end{equation}
	where $\mathbf{n}_{i} \in \mathbb{C}^{M}$ denotes the noise plus interference vector in cell $i$, modeled as
	$\mathbf{n}_{i} \sim \mathcal{CN}(\mathbf{0},\sigma^{2}\mathbf{I})$, with $\sigma^{2}=\sigma_n^2+\sigma_I^2$ being the noise plus interference power and $\mathbf{0}$ is all-zero vector. $\sigma_n^2$ is the power of white Gaussian noise and $\sigma_I^2$ is the power of interference from other cells. Here, we assume that noise and interference are identical across all cells. Accordingly, the received signal at base station $i$ from all cells can be expressed as
	\begin{equation}
		\mathbf{y}_{i}
		= \sum_{j=1}^{B} \mathbf{H}_{i,j}^{\mathrm{H}}
		\left( \sum_{k=1}^{K} \mathbf{h}_{j,k} a_{j,k} s_{j,k} + \mathbf{n}_{j} \right).
		\label{yi}
	\end{equation}
	
	\subsection{Distributed Optimization}
	In this part, we present the basic procedure of the distributed optimization framework in our work. The original objective function $F(\mathbf{x})$ is decomposed into local component functions $F_{i,k}(\mathbf{x})$ such that
	\begin{equation}
		F(\mathbf{x}) = \frac{1}{BK} \sum_{i=1}^{B} \sum_{k=1}^{K} F_{i,k}(\mathbf{x}).
	\end{equation}
	
	\subsubsection{Local Cumputing and Uplink Transmission}
	At iteration $\ell$, user $k$ in cell $i$ has access to its local function $F_{i,k}(\mathbf{x})$ and the current solution $\mathbf{x}^{(\ell)}$, and can therefore compute the corresponding local gradient $\nabla F_{i,k}(\mathbf{x}^{(\ell)})$, which is then encoded into the symbol $s_{i,k}$ and transmitted to the associated base station.
	\subsubsection{Sharing and Updating}
	 After receiving the uplink signals from all users in the cell, the base stations exchange information across cells. Each base station applies receive filtering to the superimposed signals and decodes the gradient of the global objective at the current solution, leading to the update followed:
	\begin{equation}
		\mathbf{x}_i^{(\ell+1)}
		= \mathbf{x}_i^{(\ell)} - \frac{\alpha}{BK} \sum_{i=1}^{B} \sum_{k=1}^{K} 
		\nabla F_{i,k}\big(\mathbf{x}_i^{(\ell)}\big),
	\end{equation}
	where $\alpha$ denotes the stepsize and $\mathbf{x}_i^{(\ell)}$ denotes the solution stored at base station $i$ in the $\ell$-th iteration.

	\subsubsection{Downlink Transmission}
    After computing the updated solution, each base
	station $i$ broadcasts $\mathbf{x}_{i}^{(\ell+1)}$ to all users in its cell via
	downlink transmission. Upon reception, every user updates its local copy of the
	optimization variable and uses $\mathbf{x}_{i}^{(\ell+1)}$ as the reference
	point for computing the next local gradient in the subsequent iteration. Under
	the standard assumption of a sufficiently reliable downlink control channel,
	the effects of downlink signaling errors and synchronization delays are
	neglected in the analysis.

	The above procedure is repeated until the sum of the absolute differences among the solutions at all base stations converges. The convergence criterion is specified as
	\begin{equation}
		\sum_{i=1}^{B} \big\|\mathbf{x}_{i}^{(\ell+1)} - \mathbf{x}_{i}^{(\ell)}\big\|_{2}^{2} \leq \epsilon,
	\end{equation}
	where $\epsilon > 0$ is a prescribed tolerance.

	\subsection{Problem Formulation}
	In this section, to ensure that the gradients used by the base stations to update the current solution accurately approximate the gradient of the original objective function, we formulate an optimization problem that minimizes the sum of the MSE across all base stations by jointly optimizing the receive beamforming vectors at the base stations, the transmit coefficients of the users, the movable-antenna positions at the base stations, and the transmit/receive polarization vectors. We first present the associated constraints and the objective function and , and then provide the complete optimization problem.

	 \subsubsection{Power Budget}
	 Due to the finite transmit power budget available at each user terminal, the
	 corresponding transmit coefficient $a_{i,k}$ must satisfy the following
	 per-user power constraint so that the instantaneous transmit power does not
	 exceed the maximum allowable level:
	 \begin{equation}
	 	|a_{i,k}|^{2} \leq P , \quad \forall i \in \mathcal{B},\ \forall k \in \mathcal{K},
		\label{powerconstraint}
	 \end{equation}
	 where $P$ is the power budget of each user.
	 
	 \subsubsection{Polarized Constraint}
	 As illustrated in Fig.~1, each BS is equipped with a planar antenna array, where
	 each MA is composed of two orthogonally polarized
	 radiating elements. For every MA, each polarization element is connected to a
	 dedicated RF phase shifter, so that the phase of
	 the incident signal can be independently adjusted. The resulting polarization
	 response of the array can therefore be modeled by the polarization vectors in
	 \eqref{varpi} and \eqref{mik}. Since the polarization units are
	 passive devices and thus cannot provide power amplification, the associated
	 complex weights must be properly normalized, leading to the following
	 constraints:
	 \begin{equation}
	 	|\boldsymbol{\varpi}_i(n)| = 1, \quad \forall i \in \mathcal{B},\forall n\in\{1,2\}
		\label{pconstraintvarpi}
	 \end{equation}

	 \begin{equation}
	 	|\mathbf{m}_{i,k}(n)| = 1, \quad \forall i \in \mathcal{B},\ \forall k \in \mathcal{K}, \forall n\in\{1,2\},
		\label{pconstraintm}
	 \end{equation}
	 where $\mathbf{m}_{i,k}(n)$ and $\boldsymbol{\varpi}_i(n)$ are the $n$-th entry of $\mathbf{m}_{i,k}$ and $\boldsymbol{\varpi}_i$ respectively.
	  \subsubsection{Antenna Position}
	 Due to the finite aperture size of the planar region, each MA can only move
	 within a limited area whose diameter is on the order of a few wavelengths.
	 Assuming that all BSs are equipped with planar arrays of identical physical
	 dimensions, the feasible position of the $i$-th MA is confined to the same
	 aperture region, which can be expressed as
	 \begin{equation}
	 	\tilde{\mathbf r}_i \in \mathcal{A}, \quad \forall i \in \mathcal{B},
		\label{regionconstraint}
	 \end{equation}

	 \begin{figure*}[!t]
		\vspace{-0.5\baselineskip} 
		\begin{equation}
		\begin{aligned}
		&\mathrm{MSE}(\mathbf{W},\mathbf{a},\tilde{\mathbf{r}},\mathbf{m},\boldsymbol{\varpi})
		= E\left\{ \sum_{i=1}^{B} \left| \sum_{b=1}^{B} \sum_{k=1}^{K} \frac{s_{b,k}}{BK} - y_i \right|^{2} \right\} 
		\overset{(a)}{=} E\left\{ \sum_{i=1}^{B} \left| \sum_{b=1}^{B} \sum_{k=1}^{K} \frac{s_{b,k}}{BK}
		- \sum_{j=1}^{B} \mathbf{w}_i^{\mathrm{H}} \mathbf{H}_{i,j}^{\mathrm{H}} \sum_{k=1}^{K} \mathbf{h_{j,k}} a_{j,k} s_{j,k}
		- \sum_{j=1}^{B} \mathbf{w}_i^{\mathrm{H}} n_j \right|^{2} \right\} \\
		&= \frac{1}{K}
		+ \sum_{i=1}^{B} \sum_{j=1}^{B} \sum_{k=1}^{K}
		\left\| a_{j,k} \mathbf{w}_i^{\mathrm{H}} \mathbf{H}_{i,j}^{\mathrm{H}} \mathbf{h_{j,k}} \right\|^{2}
		+ \sum_{i=1}^{B} B \sigma^{2} \left\| \mathbf{w}_i \right\|^{2} 
		- \frac{1}{BK} \sum_{i=1}^{B} \sum_{j=1}^{B} \sum_{k=1}^{K}
		a_{j,k}^{\mathrm{H}} \mathbf{h_{j,k}^{\mathrm{H}}} \mathbf{H}_{i,j} \mathbf{w}_i
		- \frac{1}{BK} \sum_{i=1}^{B} \sum_{j=1}^{B} \sum_{k=1}^{K}
		\mathbf{w}_i^{\mathrm{H}} \mathbf{H}_{i,j}^{\mathrm{H}} \mathbf{h_{j,k}} a_{j,k}.
		\end{aligned}
		\label{MSE}
		\end{equation}
		\vspace{-0.3\baselineskip} 
		\hrule
		\vspace{-0.5\baselineskip} 
		\end{figure*}

	 where
	 $\mathcal{A}$ represents the corresponding movement region.
	 
	 Moreover, the spatial correlation of the small-scale fading between antenna
	 elements can be characterized by a zeroth-order Bessel function of the first
	 kind. In practice, when the inter-element spacing exceeds approximately
	 half a wavelength, the channel fading between different antenna elements
	 can be reasonably assumed to be uncorrelated. Hence, the antenna positions
	 should satisfy
	 \begin{equation}
	 	\bigl\lvert \mathbf r_{i,m} - \mathbf r_{i,n} \bigr\rvert
	 	\ge \frac{\lambda}{2},
	 	\quad \forall i \in \mathcal{B},\ \forall m \ne n,
		\label{distanceconstraint}
	 \end{equation}
	 
	 \subsubsection{Mean Squared Error}
	 The estimation error of the proposed distributed AirComp system is quantified by the sum of the squared norms of the differences between the receive-filtered signals at each BS and the average of the symbols transmitted by all users which is given in \eqref{MSE}, 
	 where $\mathbf{a}=[\mathbf{a}_1^\mathrm{T},\cdots,\mathbf{a}_B^\mathrm{T}]^\mathrm{T} \in \mathbb{C}^{KB}$ is the collection of transmit coefficients of users with $\mathbf{a}_i=[a_{i,1},\cdots,a_{i,K}]^\mathrm{T}$, $
	 \boldsymbol{m}=\left[ \boldsymbol{m}_{1,1}^{\mathrm{T}},\cdots ,\boldsymbol{m}_{1,K}^{\mathrm{T}},\cdots ,\boldsymbol{m}_{B,K}^{\mathrm{T}} \right] ^\mathrm{T}\in \mathbb{C} ^{2BK}
	 $ is the collection of polarized vector $\mathbf{m}_{j,k}$, $\boldsymbol{\varpi}=[\boldsymbol{\varpi}_1^\mathrm{T},\cdots,\boldsymbol{\varpi}_B^\mathrm{T}]^\mathrm{T}\in \mathbb{C}^{2B}$ is the collection of polarized vector of BS. $\tilde{\mathbf{r}}=[\tilde{\mathbf{r}}_1^\mathrm{T},\cdots,\tilde{\mathbf{r}}_B^\mathrm{T}]^\mathrm{T}\in \mathbb{R}^{2BM}$ is the set of antenna positions of all base stations. The receive beamforming matrix of all base stations is denoted by
	 $\mathbf{W} = [\mathbf{w}_{1},\ldots,\mathbf{w}_{B}] \in \mathbb{C}^{M \times B}$,
	 where $\mathbf{w}_{i}$ denotes the receive beamforming vector of the base station in cell $i$. Step~(a) follows from the assumption that the noise terms at any two base stations have identical mean and variance.

	 \subsubsection{Overall Optimization Problem}
	 By collecting the above design variables and constraints, the overall joint
	 optimization problem can be formulated as
\begin{equation}
	\begin{aligned}
		\text{(P1)}\quad
		& \min_{\tilde{\mathbf{r}},\,\mathbf{a},\,\mathbf W,\,\boldsymbol{\varpi},\,\mathbf{m}} 
		 \mathrm{MSE} \\
		& \quad \quad \text{s.t.} \quad
		 \eqref{powerconstraint}\sim \eqref{distanceconstraint},
	\end{aligned}
	\label{P1}
\end{equation}

	 where constraint \eqref{powerconstraint} imposes the individual transmit-power budget for each
	 user, constraints \eqref{pconstraintvarpi} and \eqref{pconstraintm} specify the amplitude limitations of the
	 polarization vectors at the BSs and users, respectively, \eqref{regionconstraint} restricts the
	 movement region of each MA within the predefined planar aperture, and \eqref{distanceconstraint}
	 ensures a minimum separation distance between any two antenna elements.
	 
	 It is clear that problem (P1) is highly nonconvex, since the optimization
	 variables are strongly coupled in the objective function, especially through
	 the dependence of the channel matrices/vectors on both the antenna positions
	 and polarization vectors. In addition, constraint \eqref{distanceconstraint} is itself a nonconvex
	 constraint with respect to the antenna positions. As a result, (P1) cannot be
	 directly solved by off-the-shelf convex optimization solvers such as CVX.
	 Therefore, in this paper we develop an AO–based
	 iterative algorithm to obtain a high-quality suboptimal solution to (P1).

\section{PROPOSED SOLUTION}
	In this section, we propose an efficient AO algorithm to solve problem (P1). Specifically, by fixing a subset of variables and optimizing the remaining ones, the original problem is decomposed into several tractable subproblems. For the receive beamforming design, we directly derive its closed-form
	solution. For the transmit coefficient optimization subproblem, the optimal
	closed-form solution is obtained by exploiting the Karush--Kuhn--Tucker (KKT)
	conditions. For the antenna position optimization subproblem, a gradient
	descent--based iterative algorithm is employed to compute a suboptimal solution. For the subproblem of optimizing the user polarization vector, we develop an iterative algorithm based on SCA, where each iteration admits a closed-form solution. For the subproblem of optimizing the BS polarization vector, the BS polarization vector appears simultaneously in both the user–BS channel vector and the inter-BS channel matrix, rendering the resulting formulation highly nonconvex and involving polynomial terms of orders two, four, and six. To address this challenging structure, we propose an optimization framework that combines SDR with an SCA-based reformulation to transform the original problem into a tractable form, which is then solved using CVX.

	\subsection{Updating beamforming matrix $\mathbf{W}$}
	When the other optimization variables are fixed and only the receive beamforming vectors at all base stations are optimized, the original problem can be equivalently reformulated as the following optimization problem:
	\begin{equation}
		\begin{aligned}
			&(\mathrm{P2})\quad 
			\min_{\mathbf{W}}\;
			\mathrm{MSE}(\mathbf{W}\mid \mathbf{a},\tilde{\mathbf{r}},\mathbf{m},\boldsymbol{\varpi})
		\end{aligned}
		\label{P2}
	\end{equation}
	
	It is observed that this is an unconstrained convex optimization problem with respect to $\mathbf{W}$; hence, the optimal solution can be obtained by taking the gradient of the objective function and setting it to zero, which yields
	\begin{equation}
		\begin{aligned}
			\mathbf{w}_i^{\star}
			= \frac{1}{BK}
			&\Bigg(
			\sum_{j=1}^{B}\sum_{k=1}^{K}
			|a_{j,k}|^{2}\mathbf{H}_{i,j}^{H}\mathbf{h}_{j,k}\mathbf{h}_{j,k}^{H}\mathbf{H}_{i,j}
			+ B\sigma^{2}\mathbf{I}
			\Bigg)^{-1} \\
			&\times \sum_{j=1}^{B}\sum_{k=1}^{K}
			\mathbf{H}_{i,j}^{H}\mathbf{h}_{j,k}a_{j,k},
			\quad \forall i\in\{1,\ldots,B\}.
		\end{aligned}
		\label{wi}
	\end{equation}
	
	Intuitively, $\mathbf w_i^\star$ is an LMMSE-type linear estimator for the target aggregated signal: the vector $\sum_{j,k}\mathbf H_{i,j}^{\mathrm H}\mathbf h_{j,k}a_{j,k}$
	represents the composite spatial direction of the quantity to be recovered, while the matrix $	\Big(\sum_{j,k}\lvert a_{j,k}\rvert^2\,\mathbf H_{i,j}^{\mathrm H}\mathbf h_{j,k}\mathbf h_{j,k}^{\mathrm H}\mathbf H_{i,j}+B\sigma^2\mathbf I\Big)^{-1}$
	acts as a covariance-aware pre-whitening operator that accounts for the mutual coupling/correlation among the desired signals and regularizes against noise amplification through the term $B\sigma^2\mathbf I$. As a result, $\mathbf w_i^\star$ provides the MMSE-optimal tradeoff between decorrelating the multi-source useful signals and controlling noise enhancement, thereby improving the estimation quality of the desired aggregated quantity.

	\subsection{Updating transmit coefficients $\mathbf{a}$ of users}
	In this section, a closed-form solution for the users' transmit coefficients is derived. 
	To this end, an equivalent optimization problem that depends only on the transmit coefficients is first formulated. 
	Then, the problem is reformulated and its optimal solution is obtained via the KKT conditions. 
	When the other optimization variables are fixed, the original problem reduces to
	\begin{equation}
		\begin{aligned}
			\mathrm{(P3)}\quad &\min_{\mathbf{a}} \quad 
			\mathrm{MSE}(\mathbf{a}\mid \mathbf{W},\tilde{\mathbf{r}},\mathbf{m},\boldsymbol{\varpi})\\
			&~\mathrm{s.t}.\quad \eqref{powerconstraint}
		\end{aligned}
		\label{P3}
	\end{equation}

	To facilitate the derivation of the KKT structure of problem~(P3), the problem is first rewritten into a form that explicitly involves $\mathbf a$. 
	Define a unit vector $\mathbf e_{i,k}\in\mathbb{C}^{BK}$ as a selection vector whose $((i-1)K+k)$-th entry equals $1$ and all other entries equal $0$. 
	Then, using $a_{i,k}=\mathbf e_{i,k}^{\mathrm T}\mathbf a,\ \forall i,k$, problem~(P3) can be equivalently reformulated as:
	\begin{equation}\label{P3.1}
		\begin{aligned}
			(\mathrm{P3.1})\quad \min_{\mathbf a}\quad 
			& \mathbf a^{\mathrm H}\mathbf R\,\mathbf a-\mathbf a^{\mathrm H}\mathbf b-\mathbf b^{\mathrm H}\mathbf a \\
			\mathrm{s.t.}\quad 
			& \mathbf a^{\mathrm H}\mathbf E_{i,k}\,\mathbf a \le P,\quad \forall i,k,
		\end{aligned}
	\end{equation}
	where $\mathbf E_{i,k}=\mathbf e_{i,k}\mathbf e_{i,k}^{\mathrm T} \in \mathbb{C}^{BK\times BK}$, $\mathbf b=\sum_{i=1}^{B}\sum_{j=1}^{B}\sum_{k=1}^{K}\frac{1}{BK}c_{i,j,k}^{\mathrm H}\mathbf e_{j,k} \in\mathbb C^{BK}$, $c_{i,j,k}=\mathbf{w}_i^H \mathbf{H}_{i,j}^H \mathbf{h}_{j,k} \in \mathbb{C}$ and $\mathbf R=\sum_{i=1}^{B}\sum_{j=1}^{B}\sum_{k=1}^{K}\bigl|c_{i,j,k}\bigr|^{2}\mathbf e_{j,k}\mathbf e_{j,k}^{\mathrm T} \in\mathbb C^{BK\times BK}$ is positive semidefinite, for which problem (P3.1) is convex. Therefore, any solution satisfying the KKT conditions is globally optimal. 
	
	\noindent\hspace*{1em}\textit{Lemma 1.}
	The optimal solution of problem (P3.1) is given as followed:
	\begin{equation}
	a_{i,k}^\star=
	\begin{cases}
	\dfrac{b_{i,k}}{r_{i,k}}, & |b_{i,k}|\le \sqrt{P}\, r_{i,k},\\[6pt]
	\sqrt{P}\,\dfrac{b_{i,k}}{|b_{i,k}|}, & |b_{i,k}|> \sqrt{P}\, r_{i,k},
	\end{cases}
	\quad \forall\, i\in\mathcal B,\ \forall\, k\in\mathcal K,
	\label{aik}
	\end{equation}
	where $a_{i,k}$ is the $\big((i-1)K+k\big)$-th entry of $\mathbf a$,
	$b_{i,k}$ is the $\big((i-1)K+k\big)$-th entry of $\mathbf b$, and 
	$r_{i,k}$ denotes the $\big((i-1)K+k\big)$-th diagonal entry of $\mathbf R$.\\
\noindent\hspace*{1em}\textit{Proof:} Please refer to Appendix A.

	Intuitively, $\mathbf{b}$ represents the equivalent channel from the user to the base station. The resulting optimal structure of $a_{i,k}$ coincides with that of maximum ratio transmission (MRT), i.e., $a_{i,k}$ is aligned with the channel between the user and the corresponding base station.

	\subsection{Updating polarized vector $\mathbf{m}$ of users}
	In this section, we develop an iterative algorithm based on SCA for the polarization-vector optimization subproblem of each user, and derive a closed-form solution at each iteration. Fixing the other optimization variables and substituting \eqref{13c} and \eqref{18b} into the MSE expression, problem (P1) can be equivalently rewritten as
	\begin{align}
		\mathrm{(P3)}~\min_{\mathbf{m}}\;& \sum_{i=1}^{B}\sum_{j=1}^{B}\sum_{k=1}^{K}
		\left(\left|\boldsymbol{\iota}_{i,j,k}^{H}\mathbf{m}_{j,k}\right|^{2}
		-\frac{2}{BK}\Re\left\{\boldsymbol{\iota}_{i,j,k}^{H}\mathbf{m}_{j,k}\right\}\right) \label{P30_eqv}\\
		\text{s.t. }\;& \left|\mathbf{m}_{j,k}(n)\right|^{2}=1,\ \forall\, j,k,\ \forall\, n\in\{1,2\}, \nonumber
	\end{align}
	where $\mathbf{m}_{j,k}(n)$ is the $n$-th entry of vector $\mathbf{m}_{j,k}$, $
	\boldsymbol{\iota }_{i,j,k}^{H}=a_{j,k}\boldsymbol{w}_{i}^{H}\mathbf{H}_{i,j}^{H}\mathbf{h}_{j,k}^{gen}\boldsymbol{\varpi}_{j}^{H}\mathbf{A}_{j,k}^u\,\,\in \mathbb{C} ^{1\times 2}.
	$ To facilitate the subsequent derivations, we equivalently reformulate (P3) as
	\begin{align}
		\mathrm{(P3.1)}~\min_{\mathbf{m}}\;& \bar{f}(\mathbf{m})=\mathbf{m}^{H}\mathbf{D}_{\mathbf{L}}\mathbf{m}
		-\frac{2}{BK}\Re\left\{\mathbf{m}^{\mathrm{H}}\boldsymbol{\nu}\right\} \label{P3_reform}\\
		\text{s.t. }\;& \left|\mathbf{m}_{j,k}(n)\right|^{2}=1,\ \forall\, j,k,\ \forall\, n\in\{1,2\}. \nonumber
	\end{align}
	where $\mathbf{D}_L=\mathrm{blkdiag}\{[\mathbf{L}_{1,1},\cdots,\mathbf{L}_{1,K},\cdots,\mathbf{L}_{B,K}]\} \in \mathbb{C}^{2BK \times 2BK}$, the diagonal matrix of which is $\mathbf{L}_{j,k}=\sum_{i=1}^B \boldsymbol{\iota}_{i,j,k} \boldsymbol{\iota}_{i,j,k}^H \in \mathbb{C}^{2 \times 2}$. $\mathrm{blkdiag}\{[\cdot]\}$ denotes the block-diagonal matrix constructed from $[\cdot]$. $\boldsymbol{\nu}_{j,k}=\sum_{i=1}^B \boldsymbol{\iota}_{i,j,k} \in \mathbb{C}^{2 \times 1}$ is the summation of $\boldsymbol{\iota}_{i,j,k}$ and  $\boldsymbol{\nu}=\left[ \boldsymbol{\nu}_{1,1}^{T},\cdots ,\boldsymbol{\nu}_{1,K}^{T},\cdots ,\boldsymbol{\nu}_{B,K}^{T} \right]^\mathrm{T}$ is the collection of $\boldsymbol{\nu}_{j,k}$. A suitable upper bound for the objective function of (P3.1) can be constructed as \cite{huang2019reconfigurable}
	\begin{equation}
		\begin{aligned}
		\bar{f}(\mathbf{m})
		&\leq \lambda_{\max}\!\left(\mathbf{D}_{\mathbf{L}}\right)\left\|\mathbf{m}\right\|^{2}
		+2\Re\!\left\{\mathbf{m}^{H}\!\left(\mathbf{D}_{\mathbf{L}}-\lambda_{\max}\!\left(\mathbf{D}_{\mathbf{L}}\right)\mathbf{I}\right)\mathbf{m}_{t}\right\}\\
		&\quad \quad +\mathbf{m}_{t}^{H}\!\left(\lambda_{\max}\!\left(\mathbf{D}_{\mathbf{L}}\right)\mathbf{I}-\mathbf{D}_{\mathbf{L}}\right)\mathbf{m}_{t}-\frac{2}{BK}\Re\!\left\{\mathbf{m}^{\mathrm{H}}\boldsymbol{\nu}\right\}\\
		&=f(\mathbf{m} \mid \mathbf{m}_t)
		\end{aligned}
		\label{upperm}
	\end{equation}
	where $\lambda_{\max}(\mathbf{D}_L)$ is the maximum eigenvalue of matrix $\mathbf{D}_L$ and $\mathbf{m}_t$ denotes the iterate of $\mathbf{m}$ at the $t$-th iteration. It is worth noting that, by substituting \eqref{pconstraintm} into the upper bound in \eqref{upperm}, the term $\lambda_{\max}(\mathbf{D}_{\mathbf{L}})\|\mathbf{m}\|^{2}$ becomes a constant. Consequently, under the SCA framework, the optimization problem to be solved at each iteration can be written as
	\begin{equation}
	\begin{aligned}
		\mathrm{(P3.2)}~\min_{\mathbf{m}}\;
		& \Re\left\{\mathbf{m}^{H}\left(\left(2\mathbf{D}_{\mathbf{L}}-2\lambda_{\max}\!\left(\mathbf{D}_{\mathbf{L}}\right)\mathbf{I}\right)\mathbf{m}_t
		-\frac{2}{BK}\boldsymbol{\nu}\right)\right\} \\
		~\text{s.t. }\;& \left|\mathbf{m}_{j,k}(n)\right|^{2}=1,\ \forall\, j,k,\ \forall\, n\in\{1,2\}, 
	\end{aligned}
	\label{P3.2}
	\end{equation}
	Consequently, the optimal solution to (P3.2) is given by
	\begin{equation}
		\mathbf{m}^{\star}
		=\exp\!\left(j\,\angle\!\left(2\left(\lambda_{\max}\!\left(\mathbf{D}_{\mathbf{L}}\right)\mathbf{I}-\mathbf{D}_{\mathbf{L}}\right)\mathbf{m}_{t}
		+\frac{2}{BK}\boldsymbol{\nu}\right)\right),
		\label{m*}
	\end{equation}
	where the angle operator $\angle(\cdot)$ is applied element-wise. \textbf{Algorithm~1} summarizes the iterative procedure for updating $\mathbf{m}$. It is worth noting that $f(\mathbf{m}\mid \mathbf{m}^{(t)})$ satisfies the following conditions:
	\begin{equation}
	\begin{aligned}
		&f(\mathbf{m}\mid \mathbf{m}^{(t)}) \ge \bar{f}(\mathbf{m}), \quad  \\
		&f(\mathbf{m}^{(t)}\mid \mathbf{m}^{(t)}) = \bar{f}(\mathbf{m}^{(t)}),\\
		&\nabla_{\mathbf{m}} f(\mathbf{m}^{(t)}\mid \mathbf{m}^{(t)}) = \nabla_{\mathbf{m}} \bar{f}(\mathbf{m}^{(t)}). 
	\end{aligned}
	\label{suitablecondition}
	\end{equation}
	
	Therefore, the proposed algorithm is guaranteed to yield a monotonically non-increasing objective value, and its convergence is ensured.

		\begin{algorithm}[t]
			\caption{SCA Method for (P3.2)}
			\label{alg:SCA_P3}
			\begin{algorithmic}[1]
				\STATE \textbf{Initialization:} Set $t=0$. $\mathbf{m}^{(0)}$ satisfying $\mathbf{m}^{(0)}_{j,k}(n)=1,\ \forall j,k,\ \forall n\in\{1,2\}$ denotes the iterate of $\mathbf{m}$ at the $t$-th iteration, and set tolerance $\epsilon>0$ and $T_{\max}$.
				\STATE Compute $f^{(0)} \triangleq f\!\left(\mathbf{m}^{(0)}\mid \mathbf{m}^{(0)}\right)$.
				\FOR{$t=0,1,2,\ldots, T_{\max}-1$}
				\STATE Update $\mathbf{m}^{(t+1)}$ according to \eqref{m*}.
				\STATE Compute $f^{(t+1)} \triangleq f\!\left(\mathbf{m}^{(t+1)}\mid \mathbf{m}^{(t)}\right)$.
				\IF{$\left|f^{(t+1)}-f^{(t)}\right| \le \epsilon$}
				\STATE \textbf{break}
				\ENDIF
				\ENDFOR
				\STATE \textbf{Output:} $\mathbf{m}^{\star}=\mathbf{m}^{(t+1)}$.
			\end{algorithmic}
		\end{algorithm}

	\subsection{Updating polarized vector $\boldsymbol{\varpi}$ of BS}
	With the other optimization variables fixed, by substituting (13c) and (18b) into the MSE expression, the original problem (P1) can be equivalently reformulated as
	\begin{align}
		\mathrm{(P4)}~&\min_{\boldsymbol{\varpi}}\;
		\mathrm{MSE}(\boldsymbol{\varpi} \mid \mathbf{a}, \mathbf{W},\tilde{\mathbf{r}},\mathbf{m})
		\\
		&~ \text{s.t. }\; \left|\boldsymbol{\varpi}_{i}(m)\right|^{2}=1,\ \forall\, i,\ \forall\, m \in \{1,2\},
		\nonumber
	\end{align}
	Since $\boldsymbol{\varpi}$ is tightly coupled in the objective function, the resulting problem is highly nonconvex. To tackle this issue, we sequentially optimize the polarization vector of each base station. Specifically, the terms in $\sum_{i=1}^{B}\sum_{j=1}^{B}\sum_{k=1}^{K}(\cdot)$ that contain $\boldsymbol{\varpi}_m$ can be collected as $\sum_{\substack{ j\neq m}}\sum_{k=1}^{K}(\cdot)\big|_{i=m}
	+\sum_{\substack{ i\neq m}}\sum_{k=1}^{K}(\cdot)\big|_{j=m}
	+\sum_{k=1}^{K}(\cdot)\big|_{i=j=m}.$

	For brevity, we only present the derivation of the quadratic term for the case $i=j=m$:
	\begin{align}
		&\sum_{k=1}^{K}\left\|\left(\boldsymbol{\varpi}_{m}^\mathrm{H}(\mathbf{A}_{m,m}^b)^\mathrm{H}\boldsymbol{\varpi}_{m}\right)\boldsymbol{\varpi}_{m}^\mathrm{H}\boldsymbol{\epsilon}_{m,m,k}\right\|^{2} \nonumber \\
		&\overset{(a)}{=}\sum_{k=1}^{K}\left\|\left(\boldsymbol{\varpi}_{m}^\mathrm{H}\boldsymbol{\varpi}_{m}\right)\boldsymbol{\varpi}_{m}^\mathrm{H}\boldsymbol{\epsilon}_{m,m,k}\right\|^{2} \nonumber\\
		&\overset{(b)}{=}4\sum_{k=1}^{K}\left\|\boldsymbol{\varpi}_{m}^\mathrm{H}\boldsymbol{\epsilon}_{m,m,k}\right\|^{2} \nonumber\\
		&=\boldsymbol{\varpi}_{m}^\mathrm{H}\left(4\sum_{k=1}^{K}\boldsymbol{\epsilon}_{m,m,k}\boldsymbol{\epsilon}_{m,m,k}^\mathrm{H}\right)\boldsymbol{\varpi}_{m}.
		\label{sumderi}
	\end{align}
	
		\begin{algorithm}[t]
			\caption{SDR-Based Polarization Vector Update for (P4.4)}
			\label{alg:alg2_P44_final}
			\begin{algorithmic}[1]
				\STATE \textbf{Initialization:} Choose $\{\bar{\boldsymbol{\varpi}}_{m}^{(0)}\}_{m=1}^{M}$ satisfying $\bar{\boldsymbol{\varpi}}_m(n)=1, \forall m, \forall n \in \{1,2,3\}$, and set tolerance $\epsilon>0$ and $T_{\max}$.
				\FOR{$m=1,2,\ldots,M$}
				\STATE Set inner iteration index $t=0$ and initialize $\bar{\boldsymbol{\varpi}}_{m}^{(0)}$.
				\REPEAT
				\STATE Solve (P4.4) by CVX to obtain the optimal $\bar{\mathbf{V}}_{m}^{(t+1)}$.
				\STATE Apply Gaussian randomization to recover a near-optimal $\bar{\boldsymbol{\varpi}}_{m}^{(t+1)}$ from $\bar{\mathbf{V}}_{m}^{(t+1)}$.
				\STATE Update the associated parameters and compute $h_{m}^{(t+1)}$ based on $\bar{\boldsymbol{\varpi}}_{m}^{(t+1)}$.
				\STATE $t \gets t+1$.
				\UNTIL{$\big|h_{m}^{(t)}-h_{m}^{(t-1)}\big|\le \epsilon$ \textbf{or} $t\ge T_{\max}$}
				\STATE Set $\bar{\boldsymbol{\varpi}}_{m}^{\star}\gets \bar{\boldsymbol{\varpi}}_{m}^{(t)}$.
				\STATE \textbf{Recover $\boldsymbol{\varpi}_{m}^{\star}$ from $\bar{\boldsymbol{\varpi}}_{m}^{\star}$:}
				\IF{$\bar{\boldsymbol{\varpi}}_{m}^{\star}(3)=1$}
				\STATE $\boldsymbol{\varpi}_{m}^{\star}\gets \bar{\boldsymbol{\varpi}}_{m}^{\star}(1\!:\!2)$.
				\ELSIF{$\bar{\boldsymbol{\varpi}}_{m}^{\star}(3)=-1$}
				\STATE $\boldsymbol{\varpi}_{m}^{\star}\gets -\,\bar{\boldsymbol{\varpi}}_{m}^{\star}(1\!:\!2)$.
				\ENDIF
				\ENDFOR
				\STATE \textbf{Output:} $\{\boldsymbol{\varpi}_{m}^{\star}\}_{m=1}^{M}$.
			\end{algorithmic}
		\end{algorithm}

	where $\boldsymbol{\epsilon}_{i,j,k}\triangleq\!\left(\mathbf{w}_i^\mathrm{H}\left(\mathbf{H}_{i,j}^{\mathrm{gen}}\right)^\mathrm{H}\mathbf{h}_{j,k}^{\mathrm{gen}}\right)\mathbf{A}_{j,k}\mathbf{m}_{j,k}\mathbf{a}_{j,k}\in\mathbb{C}^{2 \times 1}$. 
	Step $(a)$ follows from the fact that when $i=j=m$, the elevation and azimuth angles from BS~$m$ to itself are both $0^\circ$, and thus the polarization matrix $\mathbf{A}_{i,j}^b$ degenerates into the identity matrix. 
	Step $(b)$ is obtained by exploiting the constant-modulus property of the base-station polarization vector.

	Accordingly, the objective function terms that explicitly depend on $\boldsymbol{\varpi}_m$ can be expressed as
	\begin{equation}
		\begin{aligned}
		g(\boldsymbol{\varpi}_m)&=
		\boldsymbol{\varpi}_m^{H}\!\mathbf{R}_{m,m}\!\boldsymbol{\varpi}_m		
		+2\Re\!\left\{\left(\boldsymbol{\rho}_m^{H}+\mathbf{c}_m^{H}\right)\boldsymbol{\varpi}_m\right\}\\
		&+\sum_{\substack{ i\neq m}}\sum_{k=1}^{K}
		\left(\boldsymbol{\varpi}_m^{H}\mathbf{F}_{i,m}\boldsymbol{\varpi}_m\right)
		\left(\boldsymbol{\varpi}_m^{H}\mathbf{D}_{i,m,k}\boldsymbol{\varpi}_m\right).
		\end{aligned}
		\label{gvarpim}
	\end{equation}
	$\text{where}~\mathbf{R}_{m,m}{=}\mathbf{E}_m{+}\mathbf{E}_m^{\mathrm{H}}{+}\mathbf{D}_{m,m}{+}\mathbf{N}_m, 
	\mathbf{E}_m{=}{-}\frac{1}{BK}\sum_{i\neq m}\sum_{k=1}^{K}\\ \boldsymbol{\epsilon}_{i,m,k}\boldsymbol{\varpi}_{i}^\mathrm{H}\mathbf{A}_{i,m}^b, 
	\mathbf{D}_{m,m}{=}4\sum_{k=1}^{K}\boldsymbol{\epsilon}_{m,m,k}\boldsymbol{\epsilon}_{m,m,k}^\mathrm{H}, 
	\mathbf{N}_m{=}\\ \sum_{j\neq m}\sum_{k=1}^{K}\boldsymbol{n}_{m,j,k}\boldsymbol{n}_{m,j,k}^{\mathrm{H}} ~\text{with}~ \boldsymbol{n}_{m,j,k}{=}\mathbf{A}_{m,j}^b\boldsymbol{\varpi}_{j}\boldsymbol{\epsilon}_{m,j,k}^{\mathrm{H}}\boldsymbol{\varpi}_{j}, \\
	\boldsymbol{\rho}_m{=}{-}\frac{1}{BK}\sum_{j\neq m}\sum_{k=1}^{K}\mathbf{A}_{m,j}^b\boldsymbol{\varpi}_{j}\boldsymbol{\epsilon}_{m,j,k}^{\mathrm{H}}\boldsymbol{\varpi}_{j}, 
	\mathbf{c}_m{=}{-}\frac{2}{BK} \sum_{k=1}^{K}\\ \boldsymbol{\epsilon}_{m,m,k}, 
	\mathbf{F}_{i,m}{=}(\mathbf{A}_{i,m}^b)^{\mathrm{H}}\boldsymbol{\varpi}_{i}\boldsymbol{\varpi}_{i}^\mathrm{H}\mathbf{A}_{i,m}, 
	~\text{and}~ \mathbf{D}_{i,m,k}{=}\boldsymbol{\epsilon}_{i,m,k}\boldsymbol{\epsilon}_{i,m,k}^{\mathrm{H}}.$ 
	
	Consequently, the subproblem for optimizing the beamforming vector $\boldsymbol{\varpi}_m$ at BS $m$ is formulated as
	\begin{equation}
		\begin{aligned}
			\mathrm{(P4.1)}~\underset{\boldsymbol{\varpi}_m}{\min}\quad & g\!\left(\boldsymbol{\varpi}_m\right)\\
			\text{s.t.}\quad & \left|\boldsymbol{\varpi}_m(n)\right|^{2}=1, ~\forall n\in\{1,2\}.
		\end{aligned}
		\label{P4.1}
	\end{equation}
	
	\begin{algorithm}[t]
		\caption{Gradient-Descent-Based Antenna Location Update}
		\label{alg:GD_location}
		\begin{algorithmic}[1]
			\STATE \textbf{Initialization:} Set iteration index $t=0$, choose a feasible initial point $\tilde{\mathbf{r}}^{(0)}$, set step size $\alpha>0$, tolerance $\epsilon>0$, and maximum iteration number $T_{\max}$.
			\REPEAT
			\STATE Compute $\nabla_{\tilde{\mathbf{r}}}\mathrm{MSE}$ at $\tilde{\mathbf{r}}^{(t)}$.
			\STATE Update $\tilde{\mathbf{r}}^{(t+1)}$ according to \eqref{updater}.
			\WHILE{$\tilde{\mathbf{r}}^{(t+1)} \notin \mathcal{F}$}
			\STATE $\alpha \leftarrow 0.5\,\alpha$.
			\STATE Recompute $\tilde{\mathbf{r}}^{(t+1)}=\tilde{\mathbf{r}}^{(t)}-\alpha\,\nabla_{\mathbf{r}}\mathrm{MSE}$.
			\ENDWHILE
			\STATE $t\gets t+1$.
			\UNTIL{$\big|\mathrm{MSE}^{(t)}-\mathrm{MSE}^{(t-1)}\big|\le \epsilon$ \textbf{or} $t\ge T_{\max}$}
			\STATE \textbf{Output:} $\tilde{\mathbf{r}}^{(t)}$.
		\end{algorithmic}
	\end{algorithm}

	In \eqref{gvarpim}, the presence of fourth-order terms makes the resulting objective highly nonconvex and thus difficult to optimize directly. To obtain a tractable formulation, we apply SDR method. Specifically, an auxiliary variable $t$ is introduced to construct an augmented vector, and \eqref{gvarpim} is rewritten as an expression composed of quadratic forms. By lifting the quadratic terms to a matrix variable, we obtain an equivalent lifted representation and then derive its SDR-based convex relaxation. The main reformulation steps are summarized as follows:
	\begin{subequations}\label{eq:sdr_reformulation}
		\begin{align}
			&g(\boldsymbol{\varpi}_m) \nonumber \\
			&=\bar{\boldsymbol{\varpi}}_m^\mathbf{H}\mathbf{J}_{m,m}\bar{\boldsymbol{\varpi}}_m
			+ \sum_{i\neq m}\sum_{k=1}^{K}
			\big(\bar{\boldsymbol{\varpi}}_m^\mathrm{H}\bar{\mathbf{F}}_{i,m}\bar{\boldsymbol{\varpi}}_m\big)
			\big(\bar{\boldsymbol{\varpi}}_m^\mathrm{H}\bar{\mathbf{D}}_{i,m,k}\bar{\boldsymbol{\varpi}}_m\big),
			\label{eq:sdr_reformulation_a}\\
			&= \operatorname{tr}\!\big(\mathbf{J}_{m,m}\bar{\mathbf{V}}_m\big)
			+ \sum_{i\neq m}\sum_{k=1}^{K}
			\operatorname{tr}\!\big(\bar{\mathbf{F}}_{i,m}\bar{\mathbf{V}}_m\big)\,
			\operatorname{tr}\!\big(\bar{\mathbf{D}}_{i,m,k}\bar{\mathbf{V}}_m\big),
			\label{eq:sdr_reformulation_b}
		\end{align}
	\end{subequations}
	
	\noindent where $\mathbf{J}_{m,m}=\begin{pmatrix}
		\mathbf{R}_{m,m} & \boldsymbol{\rho}_m+\mathbf{c}_m\\
		\boldsymbol{\rho}_m^\mathbf{H}+\mathbf{c}_m^\mathbf{H} & 0
	\end{pmatrix}$, $\bar{\boldsymbol{\varpi}}_m=(\boldsymbol{\varpi}_m,t)$,
	$\bar{\mathbf{V}}_m=\bar{\boldsymbol{\varpi}}_m\bar{\boldsymbol{\varpi}}_m^\mathbf{H}$,
	$\bar{\mathbf{F}}_{i,m}=\begin{pmatrix}
		\mathbf{F}_{i,m} & 0\\
		0 & 0
	\end{pmatrix}$, and
	$\bar{\mathbf{D}}_{i,m,k}=\begin{pmatrix}
		\mathbf{D}_{i,m,k} & 0\\
		0 & 0
	\end{pmatrix}$. $\bar{\mathbf{V}}_m$ is a rank-one matrix, i.e., $\operatorname{rank}\!\left(\bar{\mathbf{V}}_m\right)=1$. By dropping the rank-one constraint, we obtain the following relaxed optimization problem:
	\begin{equation}\label{eq:sdr_relaxed_problem}
	\begin{aligned}
		\mathrm{(P4.2)}~\underset{\bar{\mathbf{V}}_m}{\min}\quad & \eqref{eq:sdr_reformulation_b}\\
		\text{s.t.}\quad & \bar{\mathbf{V}}_m(n,n)=1,\quad \forall n\in\{1,2,3\},\\
		& \bar{\mathbf{V}}_m \succeq \mathbf{0}.
	\end{aligned}
	\end{equation}
	where $\bar{\mathbf{V}}_m(n,n)$ denotes the $(n,n)$-th entry of $\bar{\mathbf{V}}_m$. Owing to the nonconvexity of \eqref{eq:sdr_reformulation_b}, problem (P4.2) remains nonconvex. By exploiting the identity $\operatorname{tr}(\mathbf{A}^\mathbf{H}\mathbf{B})=\operatorname{vec}(\mathbf{A})^\mathbf{H}\operatorname{vec}(\mathbf{B})$, (P4.2) can be equivalently rewritten as
	\begin{subequations}\label{eq:P42_vec_form}
		\begin{align}
			\mathrm{(P4.3)}\underset{\bar{\mathbf{x}}_m,\ \bar{\mathbf{V}}_m}{\min}\quad 
			& \operatorname{vec}\!\big(\mathbf{J}_{m,m}^{H}\big)^\mathbf{H}\bar{\mathbf{x}}_m
			+ \bar{\mathbf{x}}_m^\mathbf{H}\mathbf{G}_m\bar{\mathbf{x}}_m\\
			\text{s.t.}\quad 
			& \bar{\mathbf{x}}_m=\operatorname{vec}\!\big(\bar{\mathbf{V}}_m\big),\label{vecx}\\
			& \bar{\mathbf{V}}_m\succeq \mathbf{0},\label{semipositive}\\
			& \bar{\mathbf{V}}_m(n,n)=1,\quad \forall n\in\{1,2,3\}.\label{const}
		\end{align}
	\end{subequations}
	where $\mathbf{G}_m=\sum_{i \neq m}\sum_{k=1}^K \operatorname{vec}(\bar{\mathbf{F}}_{i,m}^\mathrm{H})\operatorname{vec}(\bar{\mathbf{D}}_{i,m,k}^\mathrm{H})^\mathrm{H}$ and $\bar{\mathbf{x}}_m=\operatorname{vec}(\bar{\mathbf{V}}_m)$ Since $\mathbf{G}_{m}$ is not necessarily positive semidefinite, problem (P4.3) remains nonconvex. Similar to \eqref{upperm}, an appropriate upper bound for the objective function of (P4.3) can be constructed as
	\begin{equation}
		\begin{aligned}
		\operatorname{vec}\!\big(&\mathbf{J}_{m,m}^{\mathrm{H}}\big)^{\mathrm{H}}\bar{\mathbf{x}}_{m}
		+\bar{\mathbf{x}}_{m}^{\mathrm{H}}\mathbf{G}_{m}\bar{\mathbf{x}}_{m} 
		\le
		\lambda_{\max}\!\big(\mathbf{G}_{m}\big)\,\|\bar{\mathbf{x}}_{m}\|^{2}\\
		&+\Re\!\left\{\bar{\mathbf{x}}_{m}^{\mathrm{H}}
		\Big(2\mathbf{G}_{m}\bar{\mathbf{x}}_{m,t}
		-2\lambda_{\max}\!\big(\mathbf{G}_{m}\big)\bar{\mathbf{x}}_{m,t}
		+\operatorname{vec}\!\big(\mathbf{J}_{m,m}^{\mathrm{H}}\big)\Big)
		\right\} \\
		&+\bar{\mathbf{x}}_{m,t}^{H}\!\left(\lambda_{\max}\!\big(\mathbf{G}_{m}\big)\mathbf{I}-\mathbf{G}_{m}\right)\bar{\mathbf{x}}_{m,t}.
		\end{aligned}
		\label{upperx}
	\end{equation}
	where $\bar{\mathbf{x}}_{m,t}$ denotes the value of $\bar{\mathbf{x}}_{m}$ obtained at the $t$-th iteration, and $\lambda_{\max}\!\big(\mathbf{G}_{m}\big)$ is the maximum eigenvalue of $\mathbf{G}_{m}$. Therefore, under the SCA framework, the optimization subproblem to be solved at each iteration is given by
	\begin{equation}\label{eq:SCA_subproblem_P43}
		\begin{aligned}
			&\mathrm{(P4.4 )}~\underset{\bar{\mathbf{x}}_{m},\bar{\mathbf{V}}_m}{\min}\quad 
			 h_m\!\left(\bar{\mathbf{x}}_{m}\mid \bar{\mathbf{x}}_{m,t}\right)
			= \lambda_{\max}\!\big(\mathbf{G}_{m}\big)\,\|\bar{\mathbf{x}}_{m}\|^{2} \\
			&\quad +\Re\!\left\{\bar{\mathbf{x}}_{m}^{\mathrm{H}}
			\Big(2\mathbf{G}_{m}\bar{\mathbf{x}}_{m,t}
			-2\lambda_{\max}\!\big(\mathbf{G}_{m}\big)\bar{\mathbf{x}}_{m,t}
			+\operatorname{vec}\!\big(\mathbf{J}_{m,m}^{H}\big)\Big)
			\right\} \\
			&\quad \quad \quad \quad \text{s.t.}\quad 
			 \eqref{vecx}\sim \eqref{const}.
		\end{aligned}
	\end{equation}
	
	In (P4.4), if $\lambda_{\max}(\mathbf{G}_m)$ is a positive real number, then the objective function is convex. \textbf{Lemma~1} provides the corresponding proof. 
	
	\noindent\hspace*{1em}\emph{Lemma 2.}
	The maximum eigenvalue $\lambda_{\max}(\mathbf{G}_m)$ is a nonnegative real number, i.e.,
	$\lambda_{\max}(\mathbf{G}_m)\in\mathbb{R}_{+}$.\\
	\noindent\hspace*{1em}\textit{Proof: }Please refer to Appendix A.
	
	Accordingly, (P4.4) is a convex problem and can be efficiently solved using CVX. After obtaining the optimal $\bar{\mathbf{V}}_{m}$, a near-optimal $\bar{\boldsymbol{\varpi}}_{m}$ at the current iteration can be recovered via Gaussian randomization. Note that if $\bar{\boldsymbol{\varpi}}(3)=1$, then $\boldsymbol{\varpi}^*(1\!:\!2)=\bar{\boldsymbol{\varpi}}(1\!:\!2)$ and if $\bar{\boldsymbol{\varpi}}(3)=-1$, then $\boldsymbol{\varpi}^*(1\!:\!2)=-\bar{\boldsymbol{\varpi}}(1\!:\!2)$. 
	The procedure for updating $\bar{\boldsymbol{\varpi}}_{m}$ is summarized in \textbf{Algorithm~2}.

	\subsection{Updating MA positions}
	The subproblem associated with the MA locations of all BSs is given as
	\begin{equation}\label{eq:MA_location_subproblem}
		\begin{aligned}
		\mathrm{(P5)}~	&\underset{\tilde{\mathbf{r}}}{\min}\quad \mathrm{MSE}(\tilde{\mathbf{r}}|\mathbf{W},\mathbf{a},\mathbf{m},\boldsymbol{\varpi})\\
			&\text{s.t.}\quad 
			 \mathbf{r}_{i,m}\in \mathcal{A},\ \forall i,m,\\
			& \quad \quad \left\|\mathbf{r}_{i,m}-\mathbf{r}_{i,n}\right\|_{2}^{2}\ge D_{0},\ \forall i\in\mathcal{B},\ \forall m\neq n.
		\end{aligned}
	\end{equation}
	
	Since the antenna locations are coupled in the channels, the resulting problem is highly nonconvex. To tackle this difficulty, we develop a gradient-descent-based algorithm. Specifically, we first initialize the antenna location vector with a feasible point satisfying all constraints. Starting from this point, the antenna locations are updated along the negative gradient direction. If the updated locations violate the constraints, the step size is reduced to half of its previous value. In what follows, we present the gradient of the objective function in (P5) and the corresponding update rule for the antenna locations, respectively:
		\begin{align}
		&\nabla_{\tilde{\mathbf{r}}}\,\mathrm{MSE}(\tilde{\mathbf{r}})
		\triangleq
		\left[
		\left(\frac{\partial\,\mathrm{MSE}(\tilde{\mathbf{r}})}{\partial \tilde{\mathbf{r}}(1)}\right),
		\ldots,
		\left(\frac{\partial\,\mathrm{MSE}(\tilde{\mathbf{r}})}{\partial \tilde{\mathbf{r}}(2BM)}\right)
		\right]^{T},\label{gradientr} \\
		&\tilde{\mathbf{r}}^{(t+1)}=\tilde{\mathbf{r}}^{(t)}-\alpha\,\nabla_{\tilde{\mathbf{r}}}\mathrm{MSE}(\tilde{\mathbf{r}}).\label{updater}
	\end{align}
	
	\begin{algorithm}[t]
		\caption{Alternating Optimization for Solving (P1)}
		\label{alg:AO_P1}
		\begin{algorithmic}[1]
			\STATE \textbf{Initialization:} Set iteration index $t=0$, choose feasible initial values
			$\tilde{\mathbf{r}}^{(0)}$, $\mathbf{a}^{(0)}$, $\mathbf{W}^{(0)}$, $\boldsymbol{\varpi}^{(0)}$, and $\mathbf{m}^{(0)}$ satisfying \eqref{powerconstraint} $\sim$ \eqref{distanceconstraint}.
			Set tolerance $\epsilon>0$ and maximum iteration number $T_{\max}$.
			\REPEAT
			\STATE Update $\mathbf{W}$ according to \eqref{wi}.
			\STATE Update $\mathbf{a}$ according to \eqref{aik}.
			\STATE Update $\mathbf{m}$ via \textbf{Algorithm~1}.
			\STATE Update $\boldsymbol{\varpi}$ via \textbf{Algorithm~2}.
			\STATE Update $\tilde{\mathbf{r}}$ via \textbf{Algorithm~3}.
			\STATE Compute $\mathrm{MSE}^{(t+1)}$ and set $t\gets t+1$.
			\UNTIL{$\big|\mathrm{MSE}^{(t)}-\mathrm{MSE}^{(t-1)}\big|\le \epsilon$ \textbf{or} $t\ge T_{\max}$}
			\STATE \textbf{Output:} $\tilde{\mathbf{r}}^{(t)}$, $\mathbf{a}^{(t)}$, $\mathbf{W}^{(t)}$, $\boldsymbol{\varpi}^{(t)}$, and $\mathbf{m}^{(t)}$.
		\end{algorithmic}
	\end{algorithm}

	where $\tilde{\mathbf{r}}^{(t)}$ denotes the antenna location vector at the $t$-th iteration, $\alpha$ is the step size of the gradient descent update, $\tilde{\mathbf{r}}(n)$ denotes the $n$-th element of the antenna location vector $\tilde{\mathbf{r}}$. Equation~\eqref{updater} gives the antenna-location update rule. If the updated location lies outside the feasible region, the step size is reduced by half, i.e., $\alpha \leftarrow 0.5\,\alpha$. \textbf{Algorithm~3} summarizes the overall procedure of the gradient-descent method.

	\subsection{Convergence Analysis}
	The convergence of \textbf{Algorithm~4} is analyzed as follows. First, according to~\eqref{MSE}, the MSE admits a natural lower bound, i.e., $\mathrm{MSE}\ge 0$. In Steps~3 and~4, the variables are updated by the optimal solutions to their corresponding subproblems; hence, the objective value is non-increasing after each update. In Step~5, an SCA procedure is employed. The monotonicity is guaranteed by the following inequality chain:
	\begin{equation}
	\bar{f}\!\left(\mathbf{m}^{(t+1)}\right)
	\le f\!\left(\mathbf{m}^{(t+1)} \mid \mathbf{m}^{(t)}\right)
	\le f\!\left(\mathbf{m}^{(t)} \mid \mathbf{m}^{(t)}\right)
	=\bar{f}\!\left(\mathbf{m}^{(t)}\right).
	\label{nodecrease}
	\end{equation}
	Step~6 follows a similar SCA-based update rule as in Step~5. Moreover, in the Gaussian randomization stage, the randomized candidate that minimizes the original objective in~\eqref{gvarpim} is selected; therefore, the objective value does not increase in this step. Finally, Step~7 updates $\mathbf{r}$ via a gradient-descent method, which also ensures a non-increasing MSE sequence. Collectively, \textbf{Algorithm~4} generates a non-increasing sequence of objective values. Since the MSE is lower bounded, the sequence converges, which establishes the convergence of \textbf{Algorithm~4}.

	\subsection{Complexity Analysis}
	The computational complexity for updating $\mathbf{W}$ is $\mathcal{O}\!\left(BM^{3}+B^{2}KM^{2}\right)$. The complexity of updating $\mathbf{a}$ is negligible. The complexity for updating $\mathbf{m}$ is $\mathcal{O}\!\left(8B^{3}K^{3}+4T_{1}B^{2}K^{2}\right)$, where $T_{1}$ denotes the number of iterations required by \textbf{Algorithm~1}. The complexity for updating $\boldsymbol{\varpi}$ is $\mathcal{O}\!\left(MT_{2}(N_{1}+U)\right)$, where $T_{2}$ is the number of iterations in \textbf{Algorithm~2}, $U$ is the number of randomized vectors generated in the Gaussian randomization procedure and $N_{1}$ denotes the number of iterations required by CVX to solve~(P4.4). Finally, the computational complexity of the gradient-descent update in \textbf{Algorithm~3} is approximately $\mathcal{O}\!\left(T_{3}B^{3}M^{3}(L+K)\right)$, where $T_{3}$ is the number of iterations in \textbf{Algorithm~3}. As a result, the overall complexity of \textbf{Algorithm 4} is $\mathcal{O}(T_4(BM^{3}+B^{2}KM^{2}+8B^{3}K^{3}+4T_{1}B^{2}K^{2}+MT_{2}(N_{1}+U)$\\$+T_{3}B^{3}M^{3}(L+K)))$, where $T_4$ is the total number of iterations for \textbf{Algorithm 4}.

\begin{algorithm}[t]
	\caption{Two-Stage Two-Time-Scale Optimization}
	\label{alg:two_stage_two_timescale}
	\begin{algorithmic}[1]
		\REQUIRE Sample set of channel realizations $\{(\mathbf H^{(s)},\mathbf h^{(s)})\}_{s=1}^{S}$; tolerances $\epsilon_{\mathrm{in}}>0$ and $\epsilon_{\mathrm{out}}>0$; maximum iteration numbers $T_{\mathrm{in}}$ and $T_{\mathrm{out}}$.
		\STATE \textbf{Initialization:} Initialize $\tilde{\mathbf r}^{(0)}$, $\mathbf W^{(0)}$, $\mathbf a^{(0)}$, $\mathbf m^{(0)}$  and $\boldsymbol{\varpi}^{(0)}$. Set $t_{\mathrm{in}}=0$ and $t_{\mathrm{out}}=0$.
		
		\vspace{0.2em}
		\STATE \textbf{Stage I: Inner-layer optimization under instantaneous CSI.}
		\REPEAT
		\STATE Fix $\tilde{\mathbf r}=\tilde{\mathbf r}^{(0)}$ and update $\mathbf W,\mathbf a,\mathbf m, \boldsymbol{\varpi}$  by executing steps 3 $\sim$ 6 of \textbf{Algorithm~4} under the instantaneous CSI constraint \eqref{powerconstraint} $\sim$ \eqref{pconstraintm}.
		\STATE $t_{\mathrm{in}} \leftarrow t_{\mathrm{in}}+1$.
		\UNTIL{$t_{\mathrm{in}}\ge T_{\mathrm{in}}$ \OR $\left|\mathrm{MSE}^{(t_{\mathrm{in}})}-\mathrm{MSE}^{(t_{\mathrm{in}}-1)}\right|\le \epsilon_{\mathrm{in}}$}
		\STATE Denote the converged variables as $\mathbf W^{\star}$, $\mathbf a^{\star}$, $\mathbf m^{\star}$, and $\boldsymbol{\varpi}^{\star}$.
		
		\vspace{0.2em}
		\STATE \textbf{Stage II: Outer-layer optimization under statistical CSI.}
		\REPEAT
		\STATE Fix $\mathbf W=\mathbf W^{\star}$, $\mathbf a=\mathbf a^{\star}$, $\mathbf m=\mathbf m^{\star}$, and $\boldsymbol{\varpi}=\boldsymbol{\varpi}^{\star}$.
		\STATE Form the sample-average objective
		\begin{equation*}
			\overline{\mathrm{MSE}} \triangleq \frac{1}{S}\sum_{s=1}^{S}
			\mathrm{MSE}\!\left(\mathbf H^{(s)},\mathbf h^{(s)}\right).
		\end{equation*}
		\STATE Update $\tilde{\mathbf r}$ by applying \textbf{Algorithm~3}
		\STATE $t_{\mathrm{out}} \leftarrow t_{\mathrm{out}}+1$.
		\UNTIL{$t_{\mathrm{out}}\ge T_{\mathrm{out}}$ \OR $\left|\overline{\mathrm{MSE}}^{(t_{\mathrm{out}})}-\overline{\mathrm{MSE}}^{(t_{\mathrm{out}}-1)}\right|\le \epsilon_{\mathrm{out}}$}
		
		\STATE \textbf{Output:} $\tilde{\mathbf r}^{(t_{\mathrm{out}})}$, $\mathbf W^{\star}$, $\mathbf a^{\star}$, $\mathbf m^{\star}$, and $\boldsymbol{\varpi}^{\star}$.
	\end{algorithmic}
\end{algorithm}

	\section{Statistical Channel}
	The preceding analysis, as well as most related studies, is conducted under the assumption that user mobility is sufficiently slow such that the channel state remains quasi-static. In practical deployments, however, the user motion may be non-negligible, leading to time-varying base-station-to-user channels. In this case, the MA mechanism may be unable to track the rapid channel fluctuations, which can in turn degrade the communication quality. Motivated by this, a large body of prior work has advocated optimization designs based on statistical CSI. In this section, we consider a two-time-scale optimization framework. Specifically, problem (P1) is decomposed into a long-term subproblem and a short-term subproblem: in the long-term subproblem, the MA positions at each base station are optimized based on statistical CSI, whereas in the short-term subproblem, all remaining optimization variables except for the MA positions are optimized based on instantaneous CSI. Accordingly, (P1) can be equivalently reformulated as
	\begin{equation}
		\begin{aligned}
			\mathrm{(P6)}~ \min_{\tilde{\mathbf r}} \quad & \mathbb{E}_{\mathbf H,\mathbf h}\!\left\{ \min_{\mathbf W,\mathbf a,\mathbf m,\boldsymbol {\varpi}} \mathrm{MSE} \right\} \\
			\mathrm{s.t.}\quad & \eqref{powerconstraint}\sim \eqref{distanceconstraint},
		\end{aligned}
		\label{P6}
	\end{equation}
	where $\mathbf{H}$ and $\mathbf{h}$ denote the sets composed of the inter-BS channel matrices and the BS-user channel vectors, respectively. 
	The expectation operator is approximated by the sample average over multiple channel realizations, i.e.,
	\begin{equation}
		\mathbb{E}_{\mathbf H,\mathbf h}\!\left\{ \mathrm{MSE}(\mathbf H,\mathbf h) \right\}
		\approx
		\frac{1}{S}\sum_{s=1}^{S} \mathrm{MSE}\!\left(\mathbf H^{(s)},\mathbf h^{(s)}\right),
	\end{equation}
	where $S$ denotes the number of samples, and $\left(\mathbf H^{(s)},\mathbf h^{(s)}\right)$ is the $s$-th channel realization. We propose a two-time-scale optimization algorithm to solve (P6). Specifically, we first fix the MA position and iteratively solve the inner-layer problem under instantaneous CSI. After obtaining a stationary point of the inner-layer problem, we keep the resulting variables fixed, generate multiple channel samples, and then iteratively solve the outer-layer problem under statistical CSI, thereby yielding a stationary point of $\tilde{\mathbf r}$. The overall procedure is summarized in \textbf{Algorithm~5}.
	
	\subsection{Convergence Analysis}
	Here, The convergence analysis of \textbf{Algorithm~5} is provided. It is composed of an outer-loop optimization and an inner-loop optimization. The inner-loop procedure is essentially identical to \textbf{Algorithm~4}; therefore, its convergence is guaranteed. The outer-loop optimization adopts the gradient-descent method in \textbf{Algorithm~3}, which also ensures convergence. Hence, \textbf{Algorithm~5} is guaranteed to converge.

	\subsection{Complexity Analysis}
	When $\tilde{\mathbf r}$ is fixed, the computational complexity of optimizing the remaining variables is identical to that of \textbf{Algorithm~4}. For solving the outer-loop optimization problem, the computational complexity is about $\mathcal{O}(T_3SB^3M^3(L+K))$. As a result, the overall complexity of \textbf{Algorithm 5} is $\mathcal{O}(T_4(BM^3+B^2KM^2+8B^3K^3+4T_1B^2K^2+MT_2(N_1+U))+T_3SB^3M^3(L+K))$. The definitions of all parameters are consistent with those in Section~III-G, except that $T_{3}$ denotes the number of iterations for solving the outer-layer optimization problem, and $T_{4}$ denotes the number of iterations for solving the inner-layer optimization problem.

\section{Numerical Results}
In this section, we present numerical results and provide the corresponding analysis. Unless otherwise specified, the simulation parameters are set as follows. The number of cells is $B=3$, and each cell serves users $K=8$. The minimum distance between any two BSs is set to $d_{\min}=200~\mathrm{m}$. Users in each cell are assumed to be uniformly distributed within an annular region centered at their serving base station, with inner radius $20~\mathrm{m}$ and outer radius $50~\mathrm{m}$. Each BS is equipped with $4$ MAs, and the maximum transmit power of each user is $30~\mathrm{dBm}$. The carrier frequency is set to $f=3\times 10^{9}~\mathrm{Hz}$, yielding the wavelength $\lambda=c/f$, where $c$ denotes the speed of light. The feasible moving range of the MA positions at each base station is $\left[-4\lambda,\,4\lambda\right]$. The number of multipath components is $L_t = L_r = L= 3$. The channel coefficient between user $k$ and BS $i$ is modeled as
$	\mathbf{u}_{i,k}\sim \mathcal{CN}\!\left(0,\,K_0\left(\frac{d_{u,i,k}}{d_0}\right)^{-\beta}\right),\quad \forall\, i,k$,
where $d_{u,i,k}$ denotes the distance between BS $i$ and user $k$ and $\beta=1.5$ is the path-loss exponent. Moreover, the inter-BS channel coefficient matrix is modeled as
$\boldsymbol{\Sigma}_{i,j}(1,1) \sim \mathcal{CN}\!\left(0,\,K_0\left(\frac{d_{b,i,j}}{d_0}\right)^{-\beta}\frac{r}{1+r}\right), 
\boldsymbol{\Sigma}_{i,j}(n,n) \sim \mathcal{CN}\!\left(0,\,K_0\left(\frac{d_{i,j}}{d_0}\right)^{-\beta}\frac{1}{(1+r)(L-1)}\right),~ \forall\, i,j,$
where $d_{b,i,j}$ denotes the distance between base stations $i$ and $j$, $n\in\{2,\ldots,L\}$, and $L$ is the number of channel components. Here, $K_0=-40~\mathrm{dB}$ is the path gain at the reference distance, $d_0=1~\mathrm{m}$ is the reference distance and $r=1$ is the Rician factor. The white Gaussian noise power is set to $\sigma_n^{2}=-94~\mathrm{dBm}$ and the interference power is set to $\sigma_I^2=-88~\mathrm{dBm}$.

\begin{figure}[t]
	\centering
	\vspace{1mm} 
	\includegraphics[width=0.9\columnwidth, keepaspectratio, trim=0 0 0 0, clip]{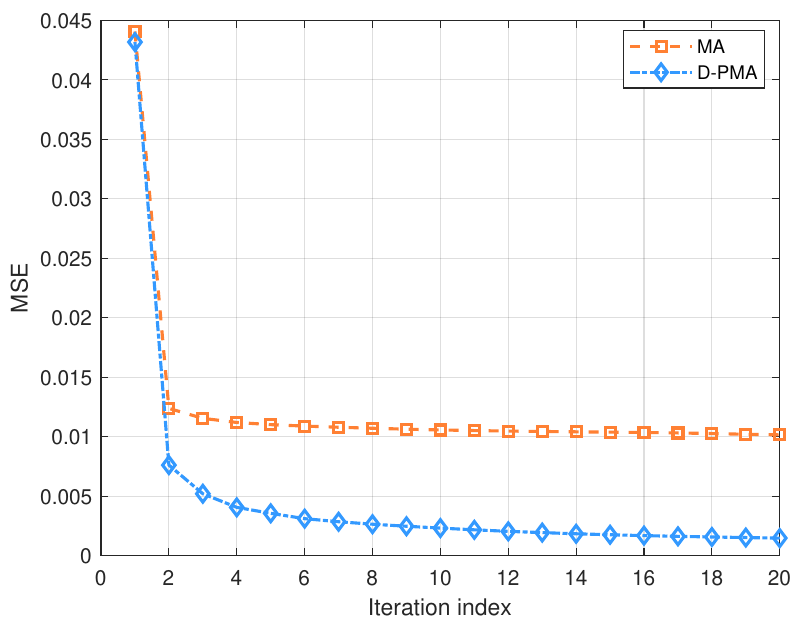}
	\caption{Iteration number versus MSE}
	\label{fig:Performance}
	\vspace{-2mm} 
\end{figure}

\begin{figure}[t]
	\centering
	\vspace{1mm} 
	\includegraphics[width=0.9\columnwidth, keepaspectratio, trim=0 0 0 0, clip]{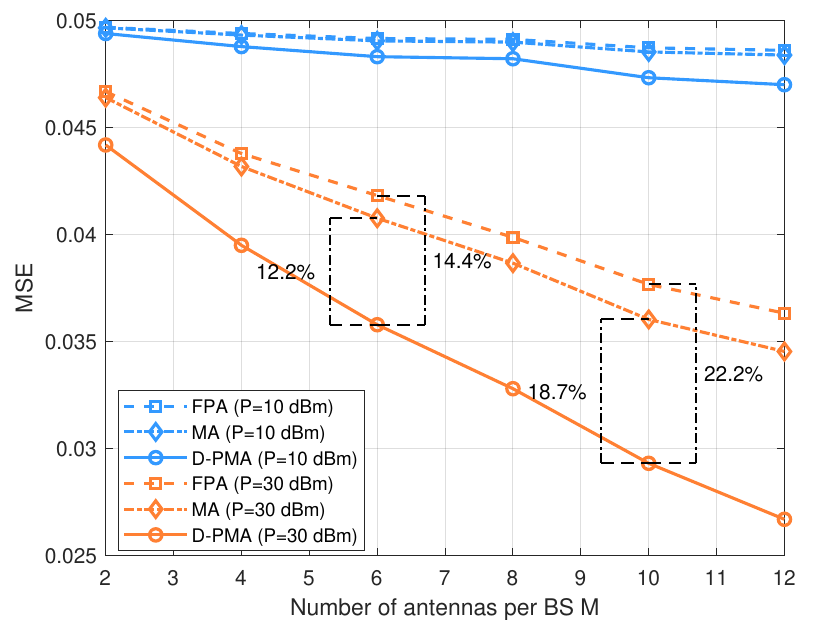}
	\caption{Number of antennas per BS versus MSE}
	\label{fig:region}
	\vspace{-2mm} 
\end{figure}

\begin{figure}[t]
	\centering
	\vspace{1mm} 
	\includegraphics[width=0.9\columnwidth, keepaspectratio, trim=0 0 0 0, clip]{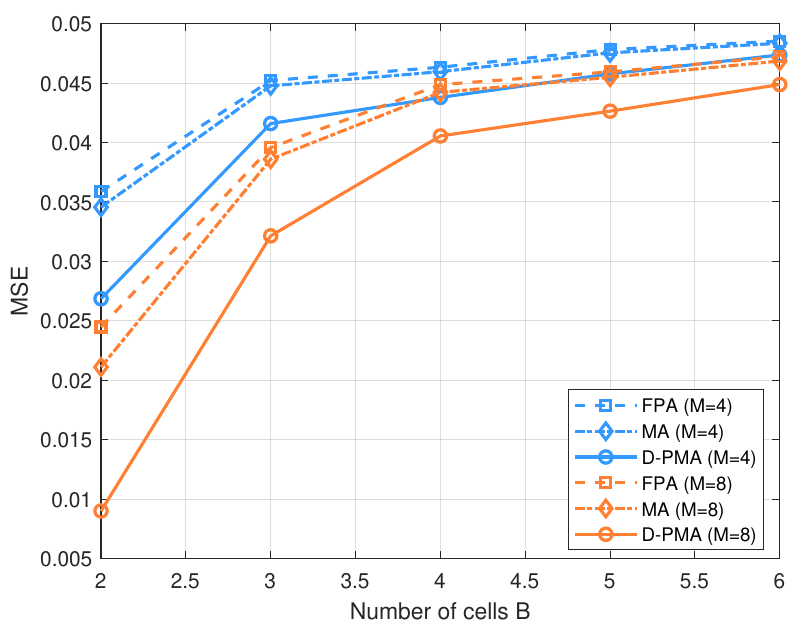}
	\caption{Number of cells versus MSE}
	\label{fig:convergence}
	\vspace{-2mm} 
\end{figure}

\begin{figure}[t]
	\centering
	\vspace{1mm} 
	\includegraphics[width=0.9\columnwidth, keepaspectratio, trim=0 0 0 0, clip]{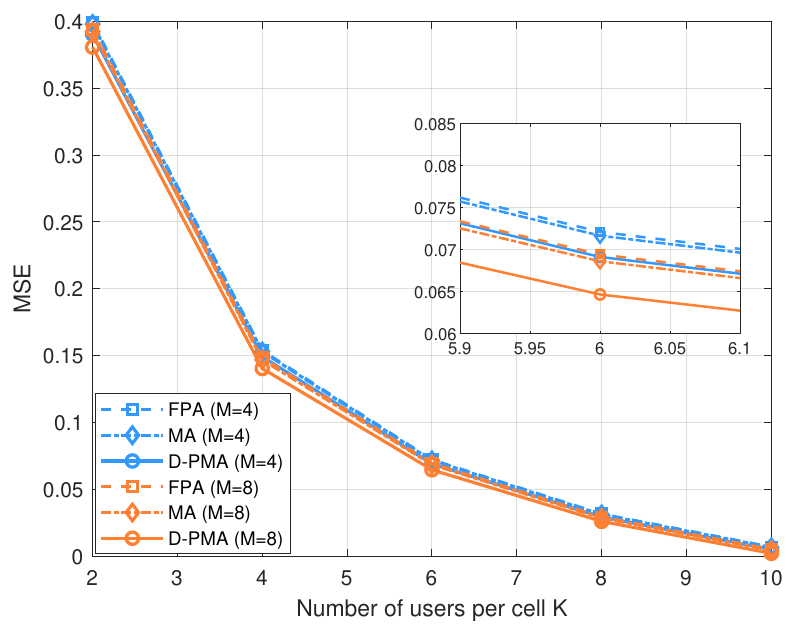}
	\caption{Number of users per cell versus MSE}
	\label{fig:convergence}
	\vspace{-2mm} 
\end{figure}

\subsection{Instantaneous Channel}
Fig.~3 illustrates the convergence behavior of the proposed D-PMA scheme in comparison with the MA scheme in the considered AirComp system. It is worth noting that the iteration index shown in Fig.~2 corresponds to the outer-loop iterations of the proposed algorithm. It can be observed that both the MA and D-PMA schemes converge within a finite number of outer iterations. Specifically, the conventional MA scheme reaches a stable solution after approximately $6$ outer iterations, whereas the proposed D-PMA scheme converges within around $12$ outer iterations. Although the D-PMA scheme requires a slightly larger number of outer iterations to converge, it consistently exhibits a faster reduction in the MSE throughout the iterative process and ultimately achieves a substantially lower steady-state MSE compared with the MA scheme. This performance gain can be attributed to the additional degrees of freedom introduced by polarization-aware optimization, which enhance signal alignment and aggregation accuracy in over-the-air computation.

Fig.~4 illustrates the MSE performance as a function of the number of antennas per BS, under different transmit power levels. It can be observed that the MSE generally decreases as $M$ increases. This trend is mainly attributed to the enhanced spatial diversity and array gain provided by a larger number of antennas, which improves the effective signal aggregation capability and mitigates the impact of noise and interference in over-the-air computation. By comparing different signal-to-noise ratio (SNR) regimes, it is observed that the MSE reduction with increasing $M$ is more pronounced in the high-SNR scenario, whereas the performance improvement becomes marginal in the low-SNR case. This is because, in the low-SNR regime, the system performance is dominated by noise, and the array gain achieved by increasing the number of antennas cannot be fully exploited. In contrast, under high-SNR conditions, the noise effect is alleviated, allowing the additional spatial degrees of freedom brought by larger $M$ to be more effectively translated into computation accuracy gains. Moreover, the proposed D-PMA framework consistently outperforms the MA and FPA schemes across all antenna configurations, with more significant gains observed in the high-SNR regime. Specifically, for $P=30$~dBm, when $M=6$, the D-PMA scheme achieves approximately $14.4\%$ and $12.2\%$ MSE reduction compared with the MA and FPA schemes, respectively. As the number of antennas increases to $M=10$, these gains further increase to about $18.7\%$ and $22.2\%$, respectively. 

\begin{figure}[t]
	\centering
	\vspace{1mm} 
	\includegraphics[width=0.9\columnwidth, keepaspectratio, trim=0 0 0 0, clip]{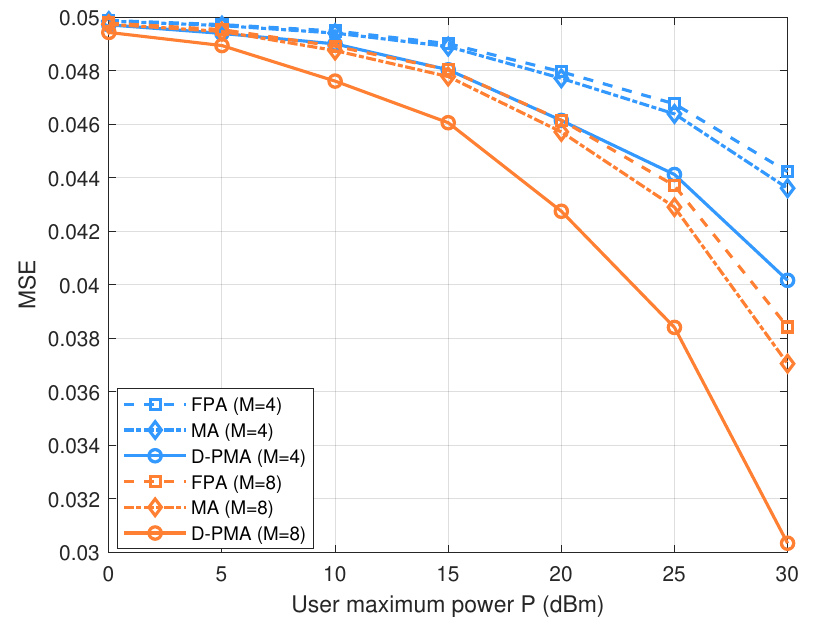}
	\caption{Transmit power versus MSE}
	\label{fig:convergence}
	\vspace{-2mm} 
\end{figure}

\begin{figure}[t]
	\centering
	\vspace{1mm} 
	\includegraphics[width=0.9\columnwidth, keepaspectratio, trim=0 0 0 0, clip]{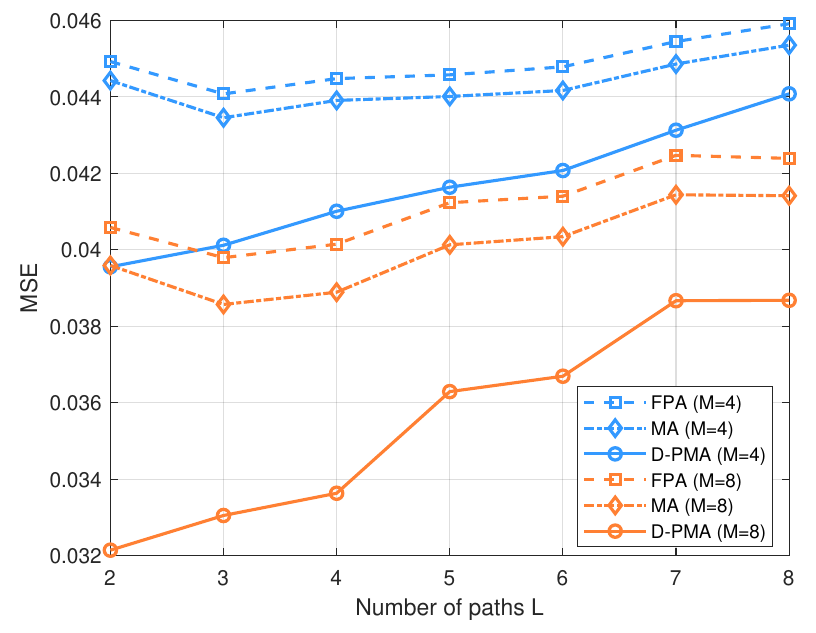}
	\caption{Number of paths versus MSE}
	\label{fig:convergence}
	\vspace{-2mm} 
\end{figure}

Fig.~5 illustrates the MSE performance as a function of the number of BSs, under different antenna configurations. It can be observed that the MSE generally increases with the number of cells for all considered schemes. This trend is expected, since increasing $B$ introduces more simultaneously transmitting cells, which leads to stronger inter-cell interference and exacerbates signal misalignment during the over-the-air aggregation process, thereby degrading the computation accuracy. It is also noteworthy that, although the MSE increases with $B$, the growth rate gradually decreases as $B$ becomes larger. From a physical perspective, when the number of cells is small, each additional cell introduces a considerable amount of new interference and aggregation distortion, resulting in a sharp increase in MSE. However, as the system scales to a larger number of cells, it gradually enters an interference-limited regime, where the marginal impact of adding one more cell becomes less significant. Consequently, the MSE growth slows down and tends to saturate. Moreover, under all considered antenna configurations, the proposed D-PMA scheme consistently achieves lower MSE than the conventional MA and FPA schemes, indicating that the additional degrees of freedom provided by polarization-aware processing can effectively mitigate the accumulation of inter-cell interference in multi-cell AirComp systems.

\begin{figure}[t]
	\centering
	\vspace{1mm} 
	\includegraphics[width=0.9\columnwidth, keepaspectratio, trim=0 0 0 0, clip]{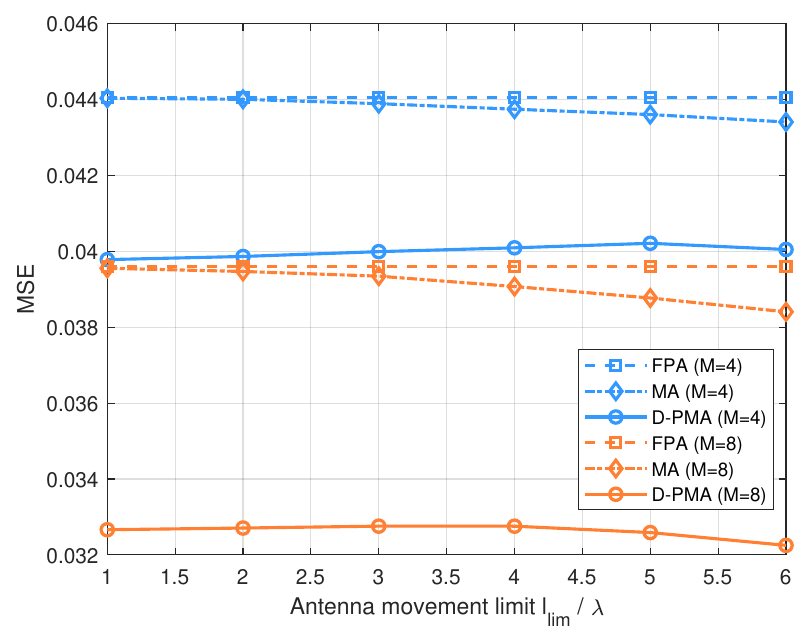}
	\caption{Antenna movement region versus MSE}
	\label{fig:convergence}
	\vspace{-2mm} 
\end{figure}

\begin{figure}[t]
	\centering
	\vspace{1mm}
	\subfloat[]{%
		\includegraphics[width=0.48\columnwidth, keepaspectratio, trim=0 0 0 0, clip]{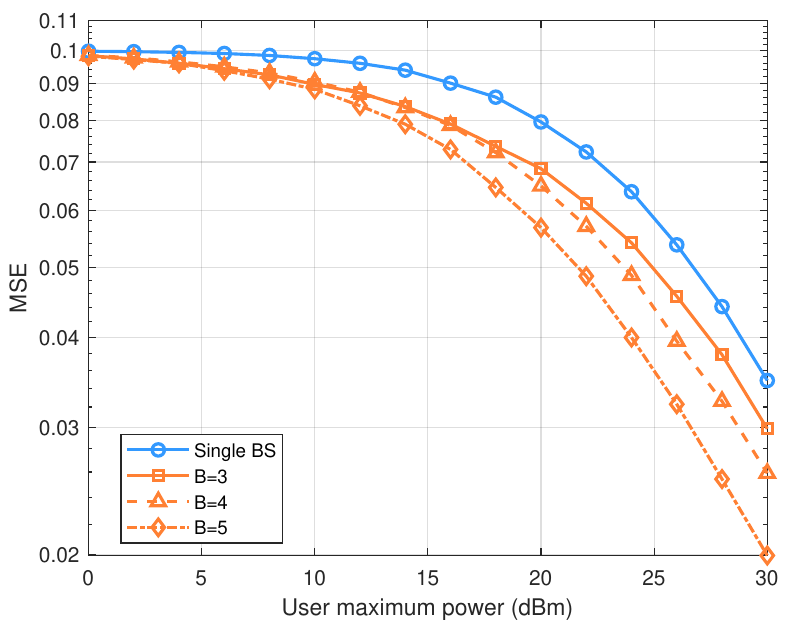}%
	}\hfill
	\subfloat[]{%
		\includegraphics[width=0.48\columnwidth, keepaspectratio, trim=0 0 0 0, clip]{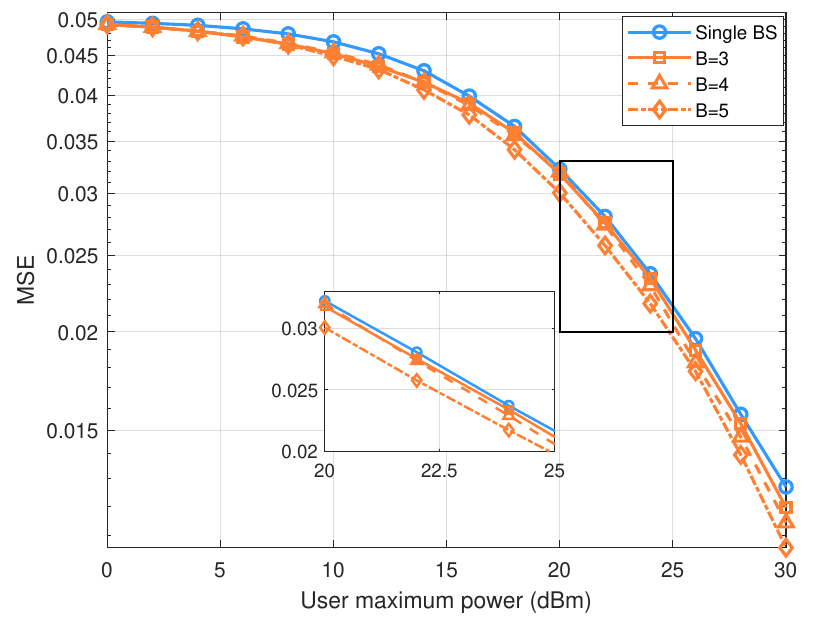}%
	}
	\caption{MSE comparison between single cell and multiple cells for D-PMA scheme with different transmit power: (a) The total number of users is 10; (b) The total number of users is 20.}
	\label{fig:stat_mse_combined}
	\vspace{-2mm}
\end{figure}

\begin{figure}[t]
	\centering
	\vspace{1mm}
	\subfloat[]{%
		\includegraphics[width=0.48\columnwidth, keepaspectratio, trim=0 0 0 0, clip]{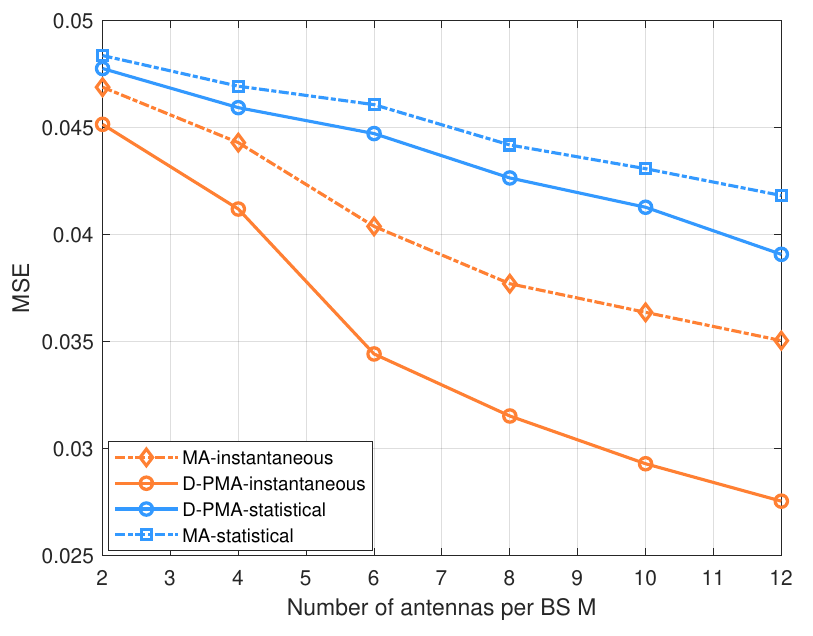}%
	}\hfill
	\subfloat[]{%
		\includegraphics[width=0.48\columnwidth, keepaspectratio, trim=0 0 0 0, clip]{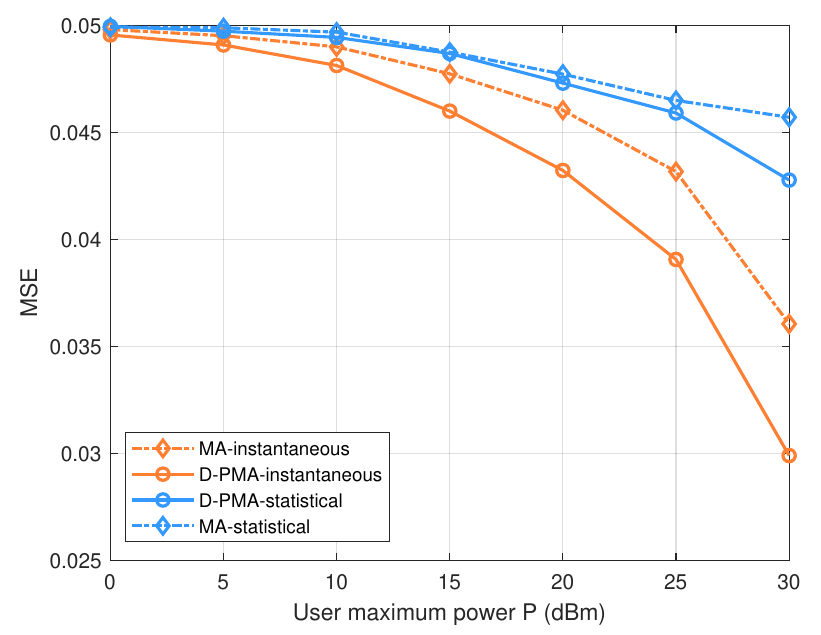}%
	}
	\caption{MSE performance under statistical channel conditions: (a) Number of antennas versus MSE; (b) Transmit power versus MSE.}
	\label{fig:stat_mse_combined}
	\vspace{-2mm}
\end{figure}

Fig.~6 and Fig.~7 depict the MSE performance of the averaged transmitted signals with respect to the number of users per cell and the maximum transmit power respectively. It can be observed that the MSE decreases monotonically as the number of users per cell increases. This behavior can be explained by the averaging nature of over-the-air computation, where the performance metric is defined as the MSE of the average of users' transmitted signals. As more users participate in the aggregation process, the random distortions caused by channel fading and noise tend to average out, thereby reducing the overall computation error. Moreover, the increased number of users enhances the concentration of the desired signal component in the superimposed waveform, leading to a more accurate estimation of the target function. Consequently, increasing $K$ results in a continuous reduction in MSE. With respect to the transmit power, the MSE decreases as the maximum user transmit power increases, which corresponds to an improvement in the SNR. A higher transmit power effectively mitigates the impact of noise on the received superimposed signal, thereby improving the computation accuracy. In addition, it is observed that the slope of the MSE curve becomes steeper in the high-SNR regime, indicating a more pronounced performance gain. This can be attributed to the fact that, once the noise effect is sufficiently suppressed, further increases in transmit power can be more efficiently translated into improved signal alignment and aggregation accuracy.

Fig.~8 and Fig.~9 show the MSE performance as a function of the number of channel paths, and the antenna movement limit respectively. It can be observed that the MSE generally increases with the number of paths. This trend can be attributed to the fact that a larger number of multipath components increases channel randomness and makes accurate signal alignment in over-the-air aggregation more challenging, thereby resulting in higher computation errors. In contrast, the MSE decreases as the antenna movement limit increases, indicating that a larger movement range provides greater spatial flexibility for adapting the antenna geometry to instantaneous channel conditions. By enabling more effective signal alignment, the aggregation accuracy is improved and the MSE is reduced. Moreover, compared with the D-PMA scheme, the MA scheme exhibits a more pronounced MSE reduction as the antenna movement range increases. This observation suggests that D-PMA already benefits from additional polarization degrees of freedom, making its performance less sensitive to geometric mobility, whereas the MA scheme relies more heavily on antenna movement to improve performance.

Fig.~10 investigates the benefits of the proposed multi-cell cooperative distributed AirComp system compared with a single-cell distributed AirComp baseline, under the setting where the total number of users and their locations are fixed. Specifically, Figs.~10(a) and 10(b) depict the MSE versus the maximum transmit power for the cases of $10$ and $20$ users, respectively, and further compare the performance under different numbers of BS. It is observed that the proposed multi-cell cooperation consistently achieves lower MSE than the single-cell system, and the performance improves monotonically as the number of cells increases. This improvement arises because the single-cell system cannot adequately accommodate users with unfavorable channel conditions, which in turn degrades the overall computation accuracy. For the case of $10$ users, when the number of cells is $B=3,4,5$, the performance gains over the single-cell baseline are approximately $11.7\%$, $23.5\%$, and $41.2\%$, respectively. Moreover, when the number of users increases to $20$, multi-cell cooperation still outperforms the single-cell scheme. However, the gain becomes noticeably smaller. This is because the total number of users approaches the saturation regime of cooperative processing across the available cells, and further improvement would require deploying more cells. Finally, we note that in Fig.~5 the total number of users increases with the number of cells, whereas in Fig.~10 the total number of users is fixed. This difference in experimental settings accounts for the seemingly different conclusions.

\subsection{Statistical Channel}
Fig.~11 compares the MSE performance of the D-PMA and MA schemes under instantaneous and statistical channel conditions. It can be observed that, under statistical channel realizations, the MA scheme exhibits relatively worse MSE performance compared with its instantaneous channel counterpart. This behavior arises because the antenna position optimization in MA is conducted based on statistical channel information, aiming at optimizing the long-term average performance rather than adapting to a specific instantaneous channel realization. Consequently, the signal alignment for a particular channel instance may be suboptimal, resulting in a larger instantaneous MSE. Nevertheless, from a long-term perspective, the MA scheme can achieve better average performance across different channel realizations. By further comparing D-PMA and MA, it is observed that although D-PMA consistently outperforms MA under both channel models, the performance gain of D-PMA over MA becomes less pronounced under statistical channel conditions. This can be attributed to the fact that, in the statistical-channel scenario, MA has already been optimized to match the channel statistics, which limits the marginal benefit that can be obtained from the additional polarization degrees of freedom. In contrast, under instantaneous channel conditions, D-PMA can more flexibly exploit polarization to adapt to channel variations, leading to more noticeable performance improvements. Moreover, under statistical channel conditions, the MSE of both MA and D-PMA decreases as the number of antennas $M$ or the maximum user transmit power $P$ increases, indicating that enlarging the antenna array or improving the transmit power enhances the signal aggregation capability in an average sense.

	\section{Conclusion} 
	In this paper, we investigate a D-PMA assisted multi-cell distributed AirComp system. By jointly optimizing the antenna positions, base-station combining beamforming matrices, polarization vectors, and user transmit coefficients, we minimize the sum of MES across all base stations. Numerical results show that, under both instantaneous and statistical CSI, the proposed D-PMA architecture consistently achieves superior performance compared with FPA and conventional MA schemes.\\

	\appendices
	\section{Derivation of Lemma 1}

	Let $\{\lambda_{i,k}\}$ denote the set of Lagrange multipliers associated with the per-element power constraints in (P3.1).
	The Lagrangian function of (P3.1) is given by
	\begin{equation}
	\begin{aligned}
	\mathcal{L}\big(\mathbf a,\{\lambda_{i,k}\}\big)
	&= \mathbf a^{\mathrm H}\!\left(\mathbf R+\sum_{i=1}^{B}\sum_{k=1}^{K}\lambda_{i,k}\mathbf E_{i,k}\right)\!\mathbf a\\
	&\quad - \mathbf b^{\mathrm H}\mathbf a - \mathbf a^{\mathrm H}\mathbf b
	- \sum_{i=1}^{B}\sum_{k=1}^{K}\lambda_{i,k} P ,
	\end{aligned}
	\end{equation}
	where $\mathbf E_{i,k}\in\mathbb R^{BK\times BK}$ is a diagonal selection matrix whose $\big((i-1)K+k\big)$-th diagonal entry equals $1$ and all other entries are $0$.
	
	Based on (P3.1) and the above Lagrangian, the KKT conditions are
	\begin{subequations}\label{eq:KKT_P31_app}
	\begin{align}
	&\left(\mathbf R+\sum_{i=1}^{B}\sum_{k=1}^{K}\lambda_{i,k}\mathbf E_{i,k}\right)\mathbf a = \mathbf b, \label{eq:KKT_P31a_app}\\
	&\lambda_{i,k} \ge 0,\quad \forall i\in\mathcal B,\ \forall k\in\mathcal K, \label{eq:KKT_P31b_app}\\
	&|a_{i,k}|^{2} \le P,\quad \forall i\in\mathcal B,\ \forall k\in\mathcal K, \label{eq:KKT_P31c_app}\\
	&\lambda_{i,k}\big(|a_{i,k}|^{2}-P\big) = 0,\quad \forall i\in\mathcal B,\ \forall k\in\mathcal K. \label{eq:KKT_P31d_app}
	\end{align}
	\end{subequations}
	
	From the stationarity condition \eqref{eq:KKT_P31a_app} and the diagonal structure of $\mathbf R$ and $\mathbf E_{i,k}$, the elements of $\mathbf a$ can be written as
	\begin{equation}\label{eq:a_elem_app}
	a_{i,k}=\frac{b_{i,k}}{r_{i,k}+\lambda_{i,k}}, \quad \forall\, i\in\mathcal B,\ \forall\, k\in\mathcal K,
	\end{equation}
	where $r_{i,k}$ denotes the $\big((i-1)K+k\big)$-th diagonal entry of $\mathbf R$.
	
	According to the complementary slackness condition \eqref{eq:KKT_P31d_app}, we consider two cases.
	
	\subsubsection{Case 1: $\lambda_{i,k}=0$}
	When $\lambda_{i,k}=0$, \eqref{eq:a_elem_app} reduces to
	\begin{equation}
	a_{i,k}=\frac{b_{i,k}}{r_{i,k}}.
	\end{equation}
	Together with the primal feasibility \eqref{eq:KKT_P31c_app}, we must have
	\begin{equation}\label{eq:case1_cond_app}
	\left|\frac{b_{i,k}}{r_{i,k}}\right|^2 \le P
	\ \Longleftrightarrow\
	|b_{i,k}|\le \sqrt{P}\, r_{i,k}.
	\end{equation}
	
	\subsubsection{Case 2: $\lambda_{i,k}>0$}
	When $\lambda_{i,k}>0$, the equality in \eqref{eq:KKT_P31d_app} must hold, i.e.,
	\begin{equation}\label{eq:active_app}
	|a_{i,k}|^2=P.
	\end{equation}
	Substituting \eqref{eq:a_elem_app} into \eqref{eq:active_app} yields
	\begin{equation}
	\left|\frac{b_{i,k}}{r_{i,k}+\lambda_{i,k}}\right|^2=P
	\ \Longleftrightarrow\
	r_{i,k}+\lambda_{i,k}=\frac{|b_{i,k}|}{\sqrt{P}},
	\end{equation}
	which gives
	\begin{equation}\label{eq:lambda_app}
	\lambda_{i,k}=\frac{|b_{i,k}|}{\sqrt{P}}-r_{i,k}.
	\end{equation}
	Using \eqref{eq:a_elem_app} and \eqref{eq:lambda_app}, we obtain
	\begin{equation}\label{eq:case2_a_app}
	a_{i,k}=\sqrt{P}\,\frac{b_{i,k}}{|b_{i,k}|}.
	\end{equation}
	Moreover, the dual feasibility \eqref{eq:KKT_P31b_app} requires $\lambda_{i,k}>0$, which implies
	\begin{equation}\label{eq:case2_cond_app}
	|b_{i,k}|>\sqrt{P}\, r_{i,k}.
	\end{equation}
	
	Combining Case~1 and Case~2, the optimal solution is finally given by
	\begin{equation}\label{eq:a_opt_app}
	a_{i,k}^\star=
	\begin{cases}
	\dfrac{b_{i,k}}{r_{i,k}}, & |b_{i,k}|\le \sqrt{P}\, r_{i,k},\\[6pt]
	\sqrt{P}\,\dfrac{b_{i,k}}{|b_{i,k}|}, & |b_{i,k}|>\sqrt{P}\, r_{i,k},
	\end{cases}
	\quad \forall\, i\in\mathcal B,\ \forall\, k\in\mathcal K,
	\end{equation}
	which completes the proof.\\

	\section{Proof of Lemma 2}\label{app:proof_lemma2}

Recall that
\begin{equation}\label{eq:Gm_def_app}
	\mathbf{G}_m \triangleq \sum_{i\neq m}\sum_{k=1}^{K}
	\operatorname{vec}\!\big(\bar{\mathbf{F}}_{i,m}^{H}\big)\,
	\operatorname{vec}\!\big(\bar{\mathbf{D}}_{i,m,k}^{H}\big)^{H}.
\end{equation}
where $N$ is an arbitrary positive integer. Define a linear operator $\mathcal{T}:\mathbb{H}^{N}\rightarrow \mathbb{H}^{N}$ over the Hermitian matrix space by
\begin{equation}\label{eq:T_def_app}
	\mathcal{T}(\mathbf{X})
	\triangleq
	\sum_{i\neq m}\sum_{k=1}^{K}
	\operatorname{tr}\!\big(\bar{\mathbf{D}}_{i,m,k}^{H}\mathbf{X}\big)\,\bar{\mathbf{F}}_{i,m}^{H},
	\qquad \mathbf{X}\in\mathbb{H}^{N}.
\end{equation}
Using $\operatorname{tr}(\mathbf{A}^{H}\mathbf{B})
=\operatorname{vec}(\mathbf{A})^{H}\operatorname{vec}(\mathbf{B})$ and the positive semidefiniteness of $\bar{\mathbf{D}}_{i,m,k}$
, we have
\begin{equation}\label{eq:trace_vec_app}
	\operatorname{tr}\!\big(\bar{\mathbf{D}}_{i,m,k}^{H}\mathbf{X}\big)
	=
	\operatorname{vec}\!\big(\bar{\mathbf{D}}_{i,m,k}^{H}\big)^{H}\operatorname{vec}(\mathbf{X}).
\end{equation}
Therefore,
\begin{align}
	\operatorname{vec}\!\big(\mathcal{T}(\mathbf{X})\big)
	&=
	\sum_{i\neq m}\sum_{k=1}^{K}
	\operatorname{tr}\!\big(\bar{\mathbf{D}}_{i,m,k}^{H}\mathbf{X}\big)\,
	\operatorname{vec}\!\big(\bar{\mathbf{F}}_{i,m}^{H}\big)\nonumber\\
	&=
	\sum_{i\neq m}\sum_{k=1}^{K}
	\operatorname{vec}\!\big(\bar{\mathbf{F}}_{i,m}^{H}\big)\,
	\operatorname{vec}\!\big(\bar{\mathbf{D}}_{i,m,k}^{H}\big)^{H}\operatorname{vec}(\mathbf{X})\nonumber\\
	&= \mathbf{G}_m\,\operatorname{vec}(\mathbf{X}), \label{eq:vec_T_app}
\end{align}
which indicates that $\mathcal{T}$ and $\mathbf{G}_m$ represent the same linear mapping under vectorization.
Hence, $\mathcal{T}$ and $\mathbf{G}_m$ share the same eigenvalues, and their spectral radii coincide
\begin{equation}\label{eq:spectral_radius_equal_app}
	\rho(\mathcal{T})=\rho(\mathbf{G}_m).
\end{equation}

Next, we show that $\mathcal{T}$ is a positive (cone-preserving) linear operator on the positive semidefinite cone
$\mathcal{K}\triangleq\{\mathbf{X}\in\mathbb{H}^{N}\mid \mathbf{X}\succeq \mathbf{0}\}$.
For any $\mathbf{X}\succeq \mathbf{0}$ and $\bar{\mathbf{D}}_{i,m,k}\succeq \mathbf{0}$, it holds that
\begin{equation}\label{eq:trace_nonneg_app}
	\operatorname{tr}\!\big(\bar{\mathbf{D}}_{i,m,k}^{H}\mathbf{X}\big)
	=\operatorname{tr}\!\big(\bar{\mathbf{D}}_{i,m,k}\mathbf{X}\big)
	=\big\|\bar{\mathbf{D}}_{i,m,k}^{\frac{1}{2}}\mathbf{X}^{\frac{1}{2}}\big\|_{F}^{2}\ge 0.
\end{equation}
Since $\operatorname{tr}(\bar{\mathbf{D}}_{i,m,k}^{H}\mathbf{X})\ge 0$ and $\bar{\mathbf{F}}_{i,m}^{H}=\bar{\mathbf{F}}_{i,m}\succeq \mathbf{0}$,
\eqref{eq:T_def_app} implies
\begin{equation}\label{eq:T_positive_app}
	\mathbf{X}\succeq \mathbf{0}\ \Longrightarrow\ \mathcal{T}(\mathbf{X})\succeq \mathbf{0},
\end{equation}
i.e., $\mathcal{T}$ is a positive linear operator.

By the Perron--Frobenius theory for positive linear operators, the spectral radius $\rho(\mathcal{T})$ is a real
nonnegative eigenvalue of $\mathcal{T}$, i.e., $\rho(\mathcal{T})\in\mathbb{R}_{+}$.
Together with \eqref{eq:spectral_radius_equal_app}, we obtain $\rho(\mathbf{G}_m)\in\mathbb{R}_{+}$.
Finally, noting that $\lambda_{\max}(\mathbf{G}_m)$ corresponds to the eigenvalue with the
largest magnitude, i.e., $\lambda_{\max}(\mathbf{G}_m)=\rho(\mathbf{G}_m)$, we conclude that
\begin{equation}\label{eq:lambda_max_nonneg_app}
	\lambda_{\max}(\mathbf{G}_m)\in\mathbb{R}_{+}.
\end{equation}
This completes the proof.


\bibliographystyle{IEEEtran}
\bibliography{ref}	

\end{document}